\documentclass[two column,aps,pra,amsmath,amssymb,preprintnumbers,superscriptaddress,longbibliography]{revtex4-2}
\usepackage{physics}
\usepackage[normalem]{ulem}
\usepackage{braket}
\usepackage{bbold}
\usepackage{amsmath, amsfonts, amssymb}
\usepackage{graphicx}
\usepackage{float}
\usepackage{lipsum}
\usepackage{subcaption}
\usepackage{float}
\usepackage{lipsum}
\usepackage{ragged2e}
\usepackage{subcaption}
\DeclareCaptionJustification{justified}{\justifying}
\usepackage[colorlinks=true,linkcolor=blue,citecolor=blue,allcolors=blue]{hyperref}
\usepackage[colorlinks=true,linkcolor=blue,citecolor=blue,allcolors=blue]{hyperref}
\usepackage{mathtools}
\usepackage{tikz}
\usepackage{natbib}
\usepackage{notes2bib}
\usepackage[ruled,vlined,linesnumbered]{algorithm2e} 

\bibnotesetup{
note-name = ,
use-sort-key = false
}
\begin{document}
\newcommand{\titleinfo}{Adaptive Tensor Network Sampling for Quantum Optimal Control}
\title{\titleinfo}

\author{Zeki Zeybek}
\email{zeki.zeybek@uni-hamburg.de}
\affiliation{The Hamburg Centre for Ultrafast Imaging, Universit{\"a}t Hamburg, Luruper Chaussee 149, 22761 Hamburg, Germany}
\affiliation{Zentrum f{\"u}r Optische Quantentechnologien, Universit{\"a}t Hamburg, Luruper Chaussee 149, 22761 Hamburg, Germany}

\author{Rick Mukherjee}
\email{rick-mukherjee@utc.edu}
\affiliation{Zentrum f{\"u}r Optische Quantentechnologien, Universit{\"a}t Hamburg, Luruper Chaussee 149, 22761 Hamburg, Germany}
\affiliation{Department of Physics \& Astronomy, University of Tennessee, Chattanooga, TN 37403, USA}
\affiliation{UTC Quantum Center, University of Tennessee, Chattanooga, TN37403, USA}

\author{Peter Schmelcher}
\email{peter.schmelcher@uni-hamburg.de}
\affiliation{The Hamburg Centre for Ultrafast Imaging, Universit{\"a}t Hamburg, Luruper Chaussee 149, 22761 Hamburg, Germany}
\affiliation{Zentrum f{\"u}r Optische Quantentechnologien, Universit{\"a}t Hamburg, Luruper Chaussee 149, 22761 Hamburg, Germany}

\begin{abstract}
Quantum optimal control (QOC) provides a systematic framework for achieving high-fidelity operations in quantum systems and plays a central role in tasks such as gate synthesis, state transfer, and pulse design. Existing QOC methods broadly fall into two categories: gradient-based and gradient-free algorithms. The associated optimization landscape is often high-dimensional, non-convex, and populated by numerous local minima, making efficient gradient-free search strategies essential. To address this, we introduce a gradient-free matrix product state/tensor train (MPS/TT) sampling heuristic for discrete quantum optimal control. In our approach, the MPS defines a score function over the space of discrete control parameters, which in turn induces a sampling distribution over candidate control sequences. This distribution is iteratively refined through selection of better performing sequences and local tensor updates to bias the search toward high-performing sequences. We evaluate the method on a range of benchmark problems, including single-qubit state transfer, Bell-pair preparation, qutrit gate implementation, and open-system population transfer. Across these tasks, the method exhibits stable convergence behavior and competitive empirical performance relative to established gradient-free baselines. These results suggest that tensor network sampling offers a viable heuristic framework for discrete quantum control.
\end{abstract}

\maketitle
	
\section{Introduction}      
Quantum optimal control concerns the task of steering a quantum system from a given initial state to a desired target state or to the implementation of a target quantum gate with maximal fidelity while respecting realistic experimental constraints such as noise, decoherence, and limited control bandwidth \cite{Ansel_2024,Koch2022}. It enables high-fidelity gate synthesis \cite{Sauvage_2022,Riaz2019}, state transfer \cite{Mukherjee_2020,Mukherjee_2020_PRL,rick3}, entanglement generation \cite{Chen_2025,Koutromanos2024_BellPair}, and time-optimal operations \cite{Dionis_2023,Wakamura_2020}. Applications span superconducting qubits \cite{Genois_2025,Abdelhafez_2020}, trapped ions \cite{Nebendahl_2009,Choi_2014}, Rydberg arrays \cite{Crescimanna_2023,Müller2016,Carrera_2025} optimizing tasks from gate compilation and calibration to robust pulse design for simulation and metrology.

The QOC problem is typically formulated as the minimization of a cost functional such as infidelity over a space of admissible control fields. The optimization landscape is often high-dimensional, non-convex, and may contain numerous local minima. This renders efficient search strategies an essential but intriguing task. Various numerical methods have been developed to tackle this challenge, broadly categorized into gradient-based \cite{Abdelhafez_2019_Gradient,Feng_2018_Gradient,Koch2022,Ansel_2024,Khane_GRAPE} and gradient-free approaches \cite{Giannelli_2022,Koch2022,Caneva_CRAB,Pagano_2024}. Gradient-based methods offer fast convergence but are hindered by barren plateaus, noise sensitivity, and reliance on accurate models. Gradient-free methods are more robust to uncertainty and experimental imperfections, yet suffer from poor sample efficiency, slow convergence, and unfavorable scaling. 

Tensor network algorithms have become versatile tools for taming the curse of dimensionality in many-body physics problems such as the celebrated density-matrix renormalization group (DMRG) algorithm \cite{Bible}. Beyond many-body physics, they find applications ranging from quantum-inspired numerical analysis \cite{Ripoll_MPS_PDE,GarciaRipoll2021quantuminspired,Bachmayr2016,Nikita_Turbulence_MPS}, probabilistic inference \cite{Bonnevie_MPS_Inference} and generative modeling \cite{Han_MPS_GenModel}, to quantum-circuit simulation \cite{Miles_MPS_Circ_DMRG,TTN_Mendl,AZ_TTN_Circ_DMRG}, and combinatorial optimization \cite{JavierEDA,Lopez_Piqueres_2023,Lopez_Piqueres_2025}. The ability to reduce parameter complexity makes them a natural fit for high-dimensional optimization. However, their application in quantum optimal control has not been fully explored.

QOC becomes computationally demanding when the number of control parameters is large since the search space grows rapidly and each objective evaluation can require a full dynamical simulation. This motivates the search for compact and adaptive search mechanisms capable of operating efficiently in high-dimensional discrete control spaces. In this work, we propose a sampling heuristic for discrete quantum optimal control based on tensor trains (TT). The central idea is to use the TT as a compressed representation of an unnormalized weight (score) over the discretized control parameter space, which in turn induces a sampling distribution over candidate control sequences. The method iteratively samples control candidates, evaluates their fidelities via dynamical simulations, and updates local TT tensors to increase the scores assigned to selected samples in high-fidelity regions. To further reduce the dimensionality, we parameterize control fields through smooth basis-function expansions that enable compact representations of continuous pulse profiles. 

\textit{Related work.\textemdash}The present approach is related to the recently introduced PROTES algorithm \cite{PROTES2023}, but differs in the objective used to update the tensor train model: PROTES updates a normalized probabilistic model, whereas TT-EDA updates unnormalized elite scores and uses the induced distribution only for sampling. In this sense, the method should not be interpreted as a probabilistic density model, but rather as a sampling heuristic that uses the TT representation to bias future samples toward promising regions of the search space. This means that selected configurations (elites) are used to increase their assigned TT weights, which in turn changes the induced sampling distribution used in subsequent iterations. TT tensors are updated using single-site alternating sweeps, in a manner reminiscent of DMRG \cite{Bible}. These distinctions are discussed in more detail in Sec.~\ref{seq:theory} and Appendix~\ref{app:protes}. To distinguish the method from PROTES, we refer to our variant as TT-EDA (Tensor Train Estimation of Distribution Algorithm).

A significant aspect of this work is to introduce TT-EDA and assess whether and how TT-EDA can be applied to standard quantum optimal control tasks, and to document its practical behavior, strengths, and limitations. To this end, we examine the performance of the proposed TT-EDA approach on a range of representative problems. These include single-qubit state transfer, Bell-pair generation in coupled spin systems, gate synthesis on a qutrit, and population transfer in an open quantum system. The observed behavior is compared against a set of commonly used gradient-free optimization baselines. Across these test cases, we find that TT-EDA can achieve performance that is competitive with established gradient-free methods within the evaluation budgets considered, with convergence behavior that depends on the problem structure and control parameterization. In single-qubit control tasks, the method is observed to recover known optimal pulse structures within a modest number of objective evaluations. For two-qubit entangling gates and qutrit gate synthesis, TT-EDA exhibits faster convergence trends while discrete gradient-free methods typically require larger evaluation budgets to reach similar fidelities. In the open-system population transfer benchmark, TT-EDA is found to reproduce the characteristic control structure associated with STIRAP-like protocols and to attain population transfer with low final infidelity under dissipative dynamics.

The paper is structured as follows. In Section \ref{seq:theory}, we introduce the optimization framework and detail the MPS/TT scoring, sampling and update scheme. We also discuss control field encoding strategies. Section \ref{seq:results} presents benchmark results across various quantum control problems, comparing TT-EDA to standard gradient-free optimizers. Finally, Section \ref{seq:outlook} concludes with a summary and outlook on future directions.

\section{Tensor sampling heuristic for optimization}
\label{seq:theory}
Probabilistic (or stochastic) optimization leverages randomization to explore complex, often non-convex search spaces and escape local minima. Rather than maintaining a single candidate solution, these methods sample candidate solutions from a stochastic search mechanism. This enables an explicit trade-off between exploration of new regions and exploitation of high-performing areas. Among these approaches, estimation of distribution algorithms (EDAs) constitute a prominent class of such methods. In EDAs, an explicit model is used to guide sampling \cite{Kochenderfer_Algorithms_for_Optimization,JavierEDA,Bergstra_Algorithms_for_Hyperparameter_Optimization}. At each iteration, candidate solutions are generated, evaluated according to the objective function, and a subset of high-performing samples is selected to update the sampling mechanism. This iterative loop biases future sampling toward regions associated with improved performance.

The connection between optimization with tensor networks and EDAs has recently been emphasized in Ref.~\cite{JavierEDA}, which interprets PROTES and related probabilistic optimization schemes within a common framework~\cite{Lopez_Piqueres_2023,Alcazar_2024}. Our approach follows the general paradigm of distribution guided optimization, but slightly differs from classical EDAs in how the search mechanism is represented and updated. Instead of fitting a globally normalized probabilistic model, we employ a tensor network parameterized sampling mechanism that generates candidate solutions through conditional sampling. Specifically, a non-negative matrix product state (MPS) defines a \emph{non-negative score (weight) function} over the high-dimensional space of discrete control sequences. Normalizing these weights implicitly induces a sampling distribution from which candidate controls can be drawn. The sampled candidates are evaluated via dynamical simulations and filtered to obtain an elite subset of high-performing configurations. The tensor network parameters are then updated through local tensor optimizations that increase the scores/weights assigned to these elite configurations. As a result, subsequent sampling becomes progressively biased toward high-fidelity regions of the search space. The method therefore retains the characteristic \emph{sample, evaluate, adapt} loop of EDAs, but it does not rely on fitting a likelihood-based density model. 

\begin{figure}[t]
\centering
    \includegraphics[width=\columnwidth]{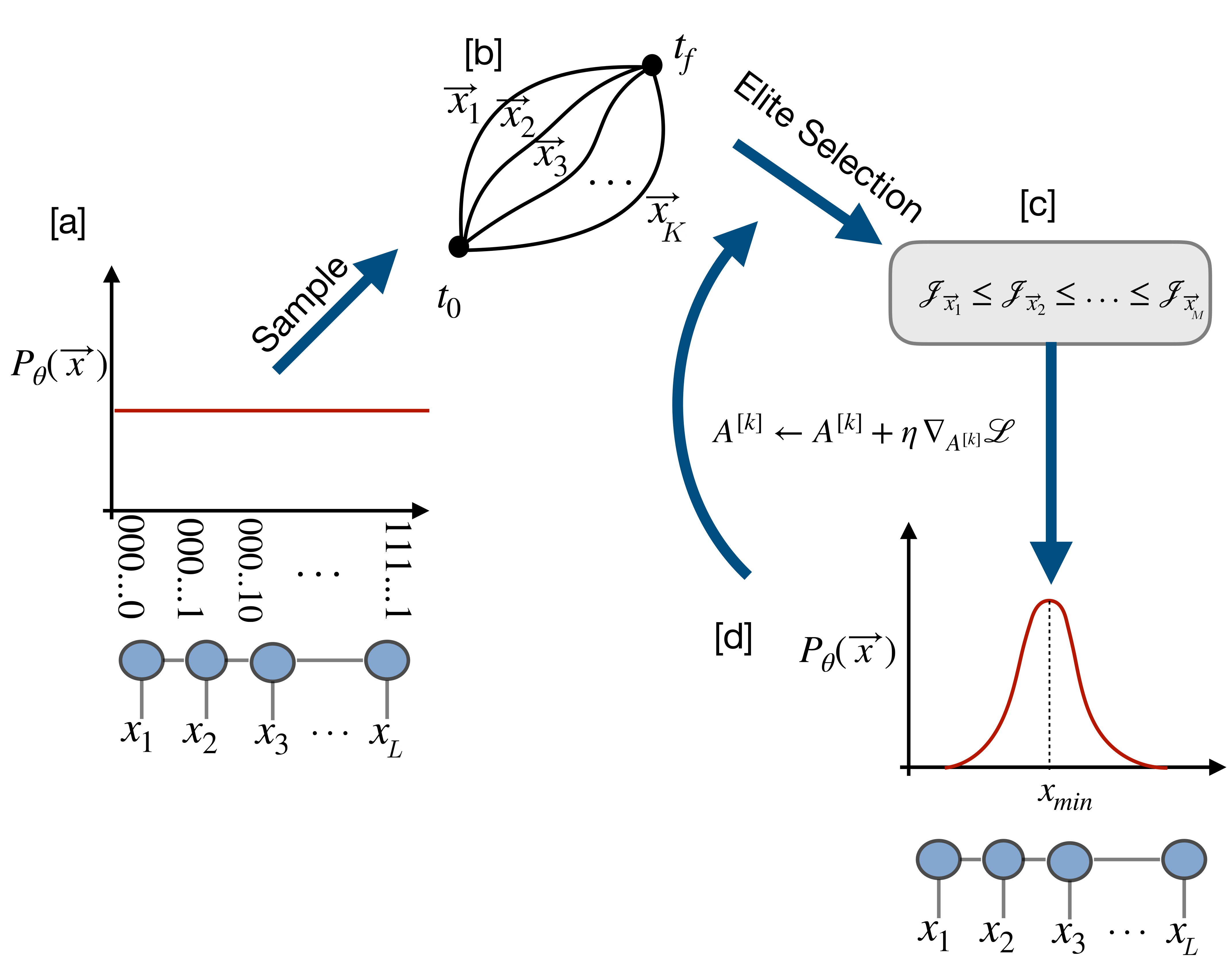}
\caption{The schematic illustration of the optimization loop is shown. (a) Initially, the uniform MPS score function defines a uniform sampling distribution over the discrete control space. (b) $K$ candidate solutions are sampled from this distribution and evaluated via dynamical simulations to obtain their performance signals. (c) A subset of $M$ high-performing samples is selected as elites, and the MPS score function parameters are updated to increase the scores of these elites, thereby biasing future sampling toward high-performing regions of the search space as shown in (d). This loop of sample, evaluate, filter and update is repeated until the optimization budget is exhausted.}
\label{fig:cycle_pic}
\end{figure}

In essence, the method should be considered as a tensor network parameterized adaptive sampling heuristic rather than a probabilistic density estimation algorithm for optimization. TT-EDA maintains an unnormalized TT/MPS score over discrete control sequences. This score induces a normalized autoregressive sampling distribution used to generate candidates, but the update step optimizes elite log-scores rather than the likelihood of a normalized density.

A key feature of the method is the use of MPS/TT representations to efficiently \emph{represent}, \emph{sample from}, and \emph{adapt} a low parameter sampling procedure over high-dimensional discrete control spaces. As also shown in Figure \ref{fig:cycle_pic}, the optimal control loop is as follows:
\begin{enumerate}
    \item \textbf{Initialization:} Define an initial score function \( S_\theta(\mathbf{x}) \) over the control parameters \( \mathbf{x} \) using MPS. 
    \item \textbf{Sampling:} Draw a set of \(\emph{K}\) candidate solutions \( \{\mathbf{x}^{(i)}\}_ {i=1}^K \) from the current distribution \( P_\theta(\mathbf{x}) \) induced by \( S_\theta(\mathbf{x}) \).
    \item \textbf{Signal:} For each sampled candidate, we extract a performance signal that reflects its quality with respect to the optimization objective. Here, the feedback signal is obtained from direct evaluations of the objective function (e.g., infidelity) for each sampled candidate solution and stored as \( \{f^{(i)}\} \).
    \item \textbf{Selection:} Identify a subset of \(\emph{M}\) high-performing (elite) configurations based on signals. This involves selecting the top \(\emph{M}\) elites.
    \item \textbf{Update:} Adjust the parameters \( \theta \) of the score function to increase the relative weights of the selected high-performing elites. This is achieved through gradient ascent on the log-scores of the elites.
    \item \textbf{Iteration:} Repeat steps 2-5 until convergence criteria are met such as reaching a maximum number of iterations or achieving satisfactory performance.
\end{enumerate}

\subsection{Scoring and Sampling with Matrix Product States}
Matrix Product States, also known as Tensor Trains, provide a compact low-rank representation of high-dimensional tensors. They can be used to parameterize the score function over a high-dimensional search space of control parameters. We consider a sequence of discrete variables \(\mathbf{x}=(x_1,\ldots,x_L)\), where \(x_k\in\{1,\ldots,d\}\).
A non-negative MPS defines the unnormalized weights/scores over the space of control sequences as,
\begin{equation}
\label{eq:mps-proposal}
S_\theta(\mathbf{x})
=
\sum_{\alpha_1,\ldots,\alpha_{L-1}}
A^{[1]}_{1,\alpha_1}(x_1)\,
A^{[2]}_{\alpha_1,\alpha_2}(x_2)\cdots
A^{[L]}_{\alpha_{L-1},1}(x_L),
\end{equation}
where \(\theta=\{A^{[k]}\}_{k=1}^L\) denotes the collection of MPS parameters.
Each local tensor \(A^{[k]}\) is associated with the variable \(x_k\): the index \(x_k\) corresponds to the physical leg of dimension \(d\), while the bond indices connect neighboring tensors with bond dimensions \(\chi\). To be explicit, boundary tensors are of shape \(1 \times d \times \chi\) and \(\chi \times d \times 1\). Although the expression in~\eqref{eq:mps-proposal} is not explicitly normalized, it can induce a normalized probability distribution via 
\begin{equation}
P_\theta(\mathbf{x}) = \frac{S_\theta(\mathbf{x})}{Z_\theta}, \quad Z_\theta = \sum_{\mathbf{x}} S_\theta(\mathbf{x}),
\end{equation}
where \(Z_\theta\) is the paritition function of the \(P_\theta\) distribution.
In practice, the partition function \(Z_\theta\) is never formed explicitly.
Instead, sampling is performed sequentially by drawing each variable from its conditional distribution given previously sampled values. For completeness, we describe the sampling \cite{vidal_sample,dolgov_sample} procedure below. The main idea is to invoke the chain rule of probability to decompose the joint distribution into a product of conditional distributions,
\begin{equation}
\label{eq:chain-rule}
P_{\theta}(x_1,x_2,\ldots,x_L)
=
P(x_1)\prod_{k=2}^L P_{\theta}(x_k \mid x_{<k}),
\end{equation}
where \(x_{<k}=(x_1,\ldots,x_{k-1})\).
Each factor in Eq.~\eqref{eq:chain-rule} is defined as, 
\begin{equation}
P_\theta(x_k \mid x_{<k})
=
\frac{S_\theta(x_{<k},x_k)}{\sum_{x_k} S_\theta(x_{<k},x_k)},
\label{eq:marginals}
\end{equation}
where \(S_\theta(x_{<k},x_k)\) can be evaluated efficiently via
MPS contractions. We initialize the left environment as \(L^{[1]}=1\) and precompute the right environments
\(R^{[k]} \in \mathbb{R}^{\chi}\) for \(k=1,\ldots,L\) by contracting the MPS tensors from right to left. At site \(k\), given the previously sampled values \(x_{<k}\), the conditional distribution over \(x_k\) is defined explicitly as the vector
\begin{equation}
\label{eq:weights}
S_\theta(x_k \mid x_{<k})
=
\sum_{\alpha_{k-1},\alpha_k}
L^{[k]}_{\alpha_{k-1}}\,
A^{[k]}_{\alpha_{k-1},\alpha_k}(x_k)\,
R^{[k]}_{\alpha_k},
\end{equation}
up to a normalization factor. After normalizing this vector, a sample \(x_k\) is drawn. After sampling, we update the left environment by absorbing \(A^{[k]}\) projected onto the sampled value \(x_k\),
\begin{equation}
\label{eq:carry-normalization}
L^{[k+1]} = L^{[k]} A^{[k]}(x_k).
\end{equation}
Repeating the procedure and caching environments for \(k=1,\ldots,L\) yields a complete sample \(\mathbf{x}=(x_1,\ldots,x_L)\).

\begin{figure}[t]
\centering
\begin{subfigure}{\columnwidth}
    \centering
    \caption{}
            \includegraphics[width=\linewidth, trim={0 0.35cm 0 0cm}, clip]{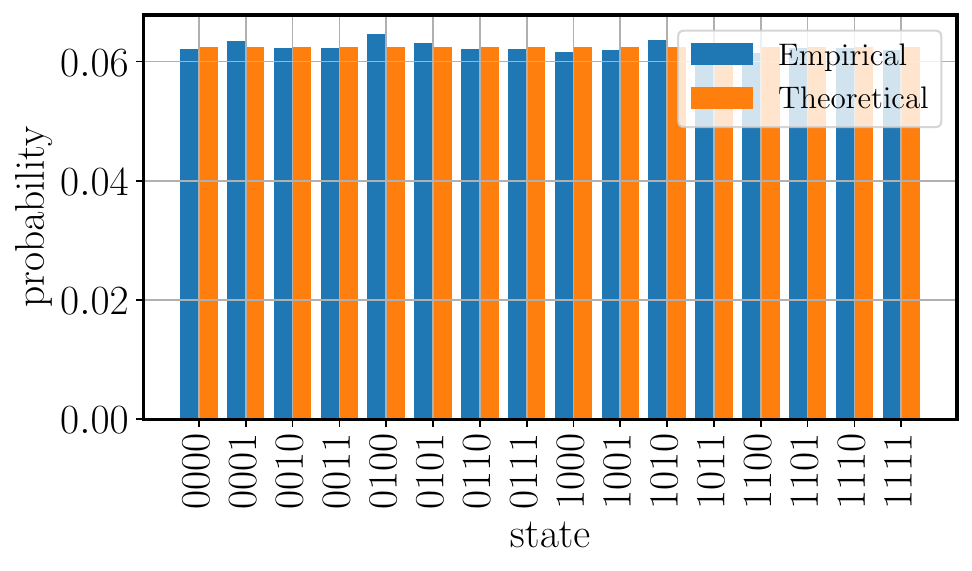}
    \label{fig:mps-sampled}
\end{subfigure}
\vspace{-1.5em} 
\begin{subfigure}{\columnwidth}
    \centering
    \caption{}
    \includegraphics[width=\linewidth, trim={0 0cm 0 0.0cm}, clip]
    {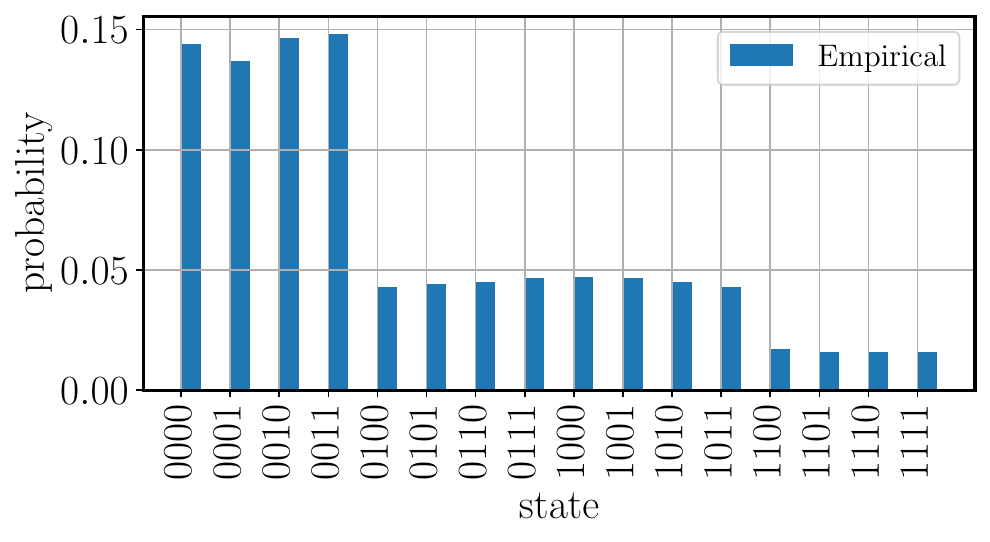}
    \label{fig:mps-updated}
\end{subfigure}
\caption{Illustration of MPS sampling and updating process. (a) Samples are drawn from a uniform distribution induced by a uniform MPS score function over bitstrings of length four. Elite samples \([0,0,0,0]\), \([0,0,0,1]\), \([0,0,1,0]\) and \([0,0,1,1]\) are selected. Here the empirical distribution is the normalized histogram obtained from samples drawn from the MPS, while the theoretical distribution is given by the exact normalized probabilities over all bitstrings computed explicitly over the full \(2^4 = 16\) configuration space. (b) After maximizing the log-scores of these elites, the MPS tensors are updated to increase their scores. Upon sampling from the distribution with the updated MPS score, samples are concentrated around the elite set.}
\label{fig:mps-proposal}
\end{figure}

\subsection{Updating Scores with Local Tensor Optimization}
Once we sample candidate optimal control sequences we evaluate their performance by means of the infidelity objective function. To do so, we simulate the system dynamics under each sampled control sequence and compute the corresponding infidelity values at the final time. Based on these evaluations, we select a subset of \(M\) elite samples with the highest performance, i.e,. \(M\) control sequences all of which generate lower infidelities than the rest. After forming this elite set, we proceed to update the MPS score function to bias future sampling toward high-performing regions of the search space. The MPS structure enables efficient updates of the scores to bias toward better samples in the search space by optimizing local tensors. Given a set of selected elite configurations $\{\mathbf{x}^{i}\}$ from the sampled batch, the local tensors \( A^{[k]} \) are updated to maximize the \textit{log-scores} objective of the following form,
\begin{equation}
\label{eq:log-score-objective}
\mathcal{L(\theta)} = \frac{1}{M}\sum_{i=1}^M \log S_\theta(\mathbf{x}^{i}).
\end{equation}
We emphasize that above in Eq.~\eqref{eq:log-score-objective}, we maximize the unnormalized weights/scores of the elite samples rather than their likelihoods. This is a key distinction from the PROTES algorithm \cite{PROTES2023}, which maximizes the likelihood of elite samples under a normalized probability model. In likelihood models, the objective would be \(\log{P_\theta(\mathbf{x}^i)}\). In contrast, our objective in Eq.~\eqref{eq:log-score-objective} is not a likelihood and does not involve normalization. Instead, it directly increases the relative scores of elite samples, which then implicitly reshapes the induced sampling distribution (see appendix \ref{app:protes}). With the above objective, we aim to achieve sampling to favor high-quality solutions. This can be achieved through gradient ascent on the tensor parameters as,
\begin{equation}
A^{[k]} \leftarrow A^{[k]} + \eta \, \nabla_{A^{[k]}} \mathcal{L},
\end{equation} 
where \( \eta \) is the learning rate. To illustrate the update procedure, let \( {\mathbf{x}}^i = ({x}^i_1,\dots,{x}^i_L) \) denote the \( i \)-th sampled configuration. For each sample \( i \) and site \( k \), we define the left \( L^i_{k} \) and right \( R^i_{k} \) environment tensors by contracting all TT tensors to the left and right of site \( k \). The gradient of the log-scores with respect to the local tensor \( A^{[k]} \) can be written compactly as
\begin{equation}
\label{eq:gradient-ascent}
\begin{aligned}
\nabla_{A^{[k]}} \mathcal{L}
&= \frac{1}{M}\sum_i \frac{\nabla_{A^{[k]}} S_\theta(\mathbf{x}^i)}{S_\theta(\mathbf{x}^i)} \\
\nabla_{A^{[k]}} S_\theta(\mathbf{x}^i)
&= L^i_k \otimes \mathbf{e}_{x^i_k} \otimes R^i_k \\
S_\theta(\mathbf{x}^i)
&= \sum_{\alpha,\beta} [L^i_k]_{1,\alpha} [A^{[k]}({x}^i_k)]_{\alpha,\beta} [R^i_k]_{\beta,1},
\end{aligned}
\end{equation}
where \( \mathbf{e}_{{x}^i_k} \in \mathbb{R}^{d} \) denotes the one-hot vector corresponding to the observed physical index \({x}^i_k \) at site \( k \) in sample \( i \) and \(A^{[k]}({x}^i_k)\) denotes the local tensor projected onto the observed physical index \({x}^i_k \). By sweeping over all sites and repeatedly applying these local updates, the MPS parameters are optimized. During each sweep, the environment tensors are updated and cached. This allows gradient contributions for all samples at a given site to be evaluated efficiently without recomputing full contractions. Since the objective in Eq.~\eqref{eq:log-score-objective} is not bounded from above, care must be taken to ensure numerical stability during optimization. Especially, gradients might blow up since environment calculations to compute gradients consists of sequential contractions of positive tensor cores. In this regard, to remove the global scaling degree of freedom of the scores and to ensure numerical stability, we renormalize MPS tensors by their Frobenius norm during optimization. This does not affect the induced conditional sampling probabilities. We also use gradient clipping as a practical safeguard against large gradient values. While not required by the formulation, this was observed to improve stability in some cases. Since our approach is a rather aggressive sampling heuristic (more details in appendix \ref{app:protes}), we incorporate random mutations to maintain exploration. After sampling from the MPS, each variable \( x_k \) is perturbed with a small probability \( \varepsilon \), replacing it with a uniformly random value from its alphabet.

Figures~\ref{fig:mps-sampled} and \ref{fig:mps-updated} illustrate the sampling and update mechanism of the MPS. Initially, samples are drawn from a uniform scoring model (Fig.~\ref{fig:mps-sampled}). In this toy example, the scores over all bitstrings of length four are initialized uniformly (with normalization applied only for sampling convenience). As elite configurations, we select \([0,0,0,0]\), \([0,0,0,1]\), \([0,0,1,0]\) and \([0,0,1,1]\). The MPS tensors are then updated to increase the relative scores of these elite configurations. This biases subsequent sampling toward these regions of the search space. This effectively leads to concentrating samples around the elite set (Fig.~\ref{fig:mps-updated}).

\subsection{Control Field Encoding} 
In QOC, the control fields \( u(t) \) are functions of time that influence the system dynamics. Encoding scheme for these control variables should be adopted to define the score function over the search space of control fields. In this work, we consider two encoding strategies: \( (i) \) direct discretization of the control fields as time series and \( (ii) \) basis-function parameterizations. 

\subsubsection{Direct Discretization: Time-Series of Control Fields}
In the direct discretization scheme, the control field is parameterized directly in the time domain by discretizing the control amplitude on a uniform temporal grid
\(t_1, t_2, \ldots, t_L.\) The control field is represented as a discrete sequence \(\mathbf{u} = (u_{t_1}, u_{t_2}, \ldots, u_{t_L}),\) where each time slice corresponds to an independent control degree of freedom. To integrate this representation into the optimization framework, each control amplitude is discretized into a finite set of admissible values via a symbol \(x_k \in \{1,\ldots,d\},  u_{t_k} = b(x_k),\)
where \( b:\{1,\ldots,d\}\to\mathbb{R} \) is a value map defining the allowed amplitude levels. The discrete control sequence \(\mathbf{x} = (x_1, x_2, \ldots, x_L)\)
thus serves as the optimization variable. The score over time-discretized control amplitudes is modeled by a tensor train as \(S_\theta(\mathbf{x})\). In this construction, the length of the MPS is equal to the number of discretized time steps. In the MPS, each site corresponds to a time step \( t_k \) and the physical leg represents the discrete set of possible \( d \) control amplitudes at that time. The mapping \( b:\{1,\ldots,d\}\to\mathbb{R} \) associates each discrete symbol \( x_k \) with a physical control amplitude. In practice, this corresponds to discretizing a continuous amplitude range into a finite set of admissible values. For example, given a control amplitude constrained to the interval
\( u(t)\in[u_{\min},u_{\max}] \), one may define a uniform discretization
\[
b(m) = u_{\min} + (m-1)\,\frac{u_{\max}-u_{\min}}{d-1}, 
\qquad m=1,\ldots,d,
\]
so that each symbol \( x_k \) selects one of \( d \) evenly spaced amplitude levels.
More generally, the map \( b \) may be chosen non-uniformly to emphasize regions of
interest or to respect experimental constraints. As a concrete example, consider a control amplitude constrained to
\(u(t)\in[-4,4]\) and a discretization with \(d=6\) levels. The mapping
\(b\) may be defined as
\(b = \{-4,\,-2.4,\,-0.8,\,0.8,\,2.4,\,4\}\).
Sampling a discrete variable \(x_k=4\) then yields the physical control
value \(u_{t_k}=b(4)=0.8\). The length of the MPS scales linearly with the number of discretized time steps \( L \). The representation can then become less efficient when very fine temporal resolutions are required to accurately capture smooth control fields. The discretization of control amplitudes introduces an inherent approximation error. This may lead to discretization artifacts and ultimately limit the achievable control fidelity.

\subsubsection{Basis Encoding}
An alternative to direct time-series encoding is to parameterize control fields using a finite set of basis functions. In this approach, the control field is represented as
\[
u(t) = \sum_{j=1}^{L} c_j \, \phi_j(t),
\]
where \( \{\phi_j(t)\} \) denotes a chosen basis and \( \mathbf{c} = (c_1,\ldots,c_L) \) are the corresponding expansion coefficients. By optimizing over basis coefficients rather than individual time samples, this representation reduces the effective dimensionality of the control space and introduces structural priors such as smoothness through the choice of basis. To incorporate basis-encoded controls into the optimization framework, each coefficient \( c_j \) is discretized into a finite set of admissible values via a symbol
\( x_j \in \{1,\ldots,d\} \) and a value map \( b:\{1,\ldots,d\}\to\mathbb{R} \), such that
\( c_j = b(x_j) \). The score over the coefficient vector \( \mathbf{x}=(x_1,\ldots,x_L) \) is modeled by a MPS \(S_\theta(\mathbf{x})\). In this encoding, each site corresponds to a basis coefficient, with a physical dimension \( d \) and bond dimension \( \chi \). We briefly describe several concrete choices of basis functions.

\textit{Piecewise-constant basis.\textemdash}
The control is expanded over \( L \) piecewise-constant basis functions,
\[
u(t)=\sum_{j=1}^{L} c_j\,\phi_j(t),\qquad 
\phi_j(t)=\mathbb{1}_{[t_{j-1},\,t_j)}(t).
\]
Here, \( \mathbb{1}_{[a,b)}(t) \) denotes the indicator function over the interval \([a,b)\). The number of coefficients \( L \) corresponds to the number of time segments.

\textit{Fourier basis.\textemdash}
The control field is expanded in a truncated Fourier series,
\[
u(t)=c_0 + \sum_{\ell=1}^{J}\big[c_{\ell}^{(c)}\cos(\ell t) + c_{\ell}^{(s)}\sin(\ell t)\big],
\]
yielding \( L=2J+1 \) coefficients. 

\textit{Spline basis.\textemdash}
The control is represented using a B-spline basis of degree \( p \) over a knot vector \( \mathbf{k} \),
\[
u(t)=\sum_{j=1}^{L} c_j\,B_{j,p}(t;\,\mathbf{k}).
\]
Here, \( B_{j,p}(t;\,\mathbf{k}) \) denotes the \( j \)-th B-spline basis function of degree \( p \). The number of coefficients \( L \) is determined by the number of basis functions.

\section{Results}
\label{seq:results}
We consider a set of QOC problems including single- and two-qubit state transfer, gate synthesis, and open-system state transfer. For each problem, the objective is to minimize a task-specific cost functional, such as
state infidelity and gate infidelity evaluated through numerical simulation of the system dynamics. To examine  performance, we benchmark our approach against a range of well-known continuous
and discrete gradient-free optimization algorithms implemented in the Nevergrad \cite{nevergrad} library.
These include population-based and evolutionary strategies as well as other
derivative-free optimizers. The abbreviations used throughout are: DE (differential evolution), PSO (particle swarm optimization), CMA (covariance matrix adaptation evolution strategy), NGO (Nevergrad
meta-optimizer), DOPO (discrete one-plus-one evolutionary algorithm), DFG
(double-fast genetic algorithm), and PDOPO (portfolio of multiple one-plus-one strategy). Hyperparameters for all optimizers are set to their default values in Nevergrad. Each benchmark is repeated over multiple independent runs with different random seeds.

\subsection{Single-Qubit Population Transfer}
In this test case, we are interested in state-to-state transfer of a two-level quantum system. We consider a two-level quantum system
whose Hamiltonian can be written as
\begin{equation}
    \label{eq: single-qubit-hamiltonian}
    {H}(t) = \frac{\Delta}{2}\sigma_z + \frac{u(t)}{2}\sigma_x,
\end{equation}
where \(\sigma_x\) and \(\sigma_z\) are the Pauli matrices, \(\Delta\) is the detuning, and \(u(t) \in \{-u_0, u_0\}\) is the control field. The goal is to transfer the system from the ground state \(\ket{0}\) to the excited state \(\ket{1}\) in minimum time \(T\) while minimizing the infidelity objective function (or figure of merit) \(\mathcal{J} = 1 - \abs{\braket{1|\psi(T)}}^2 \). In this optimal control problem, the control field is parameterized as a discrete time series on a uniform grid with discretized control amplitudes into finite levels. The MPS score is defined over the space of these discrete control amplitudes at each time step. We consider two scenarios: (i) resonant control with \(\Delta = 0\) and (ii) off-resonant control with \(\Delta \neq 0\). 

\begin{figure}[t]
    \centering
    \begin{subfigure}{\columnwidth}
        \centering
        \caption{}
        \includegraphics[width=\linewidth]{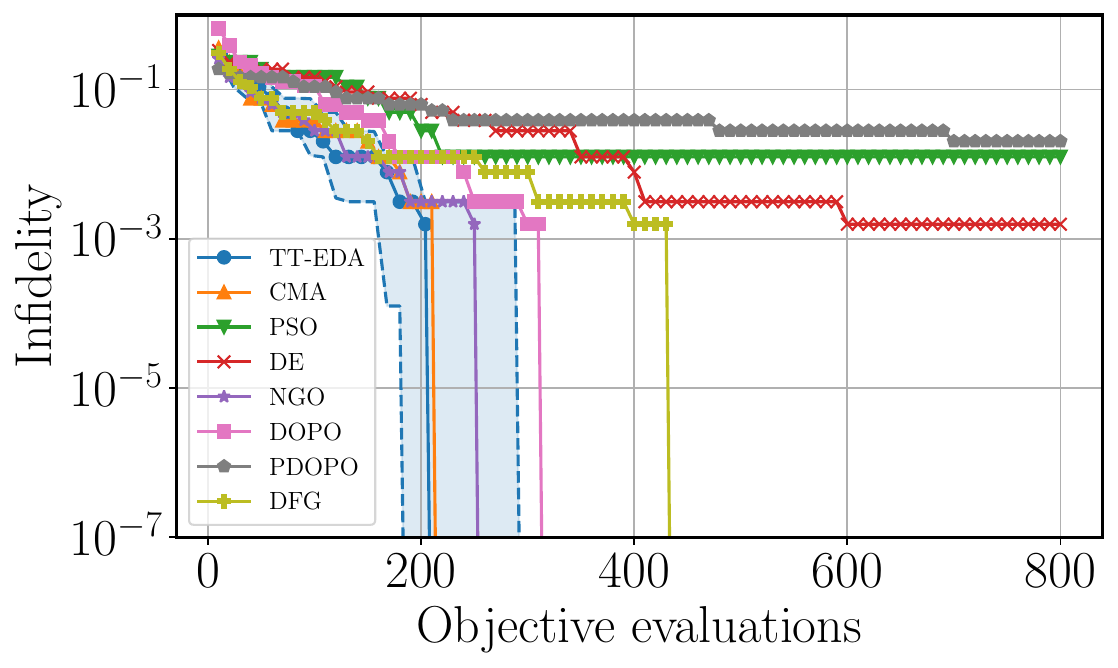}
        \label{fig:det0-convergence}
    \end{subfigure}

    \begin{subfigure}{0.48\columnwidth}
        \centering
        \caption{}
        \includegraphics[width=\linewidth]{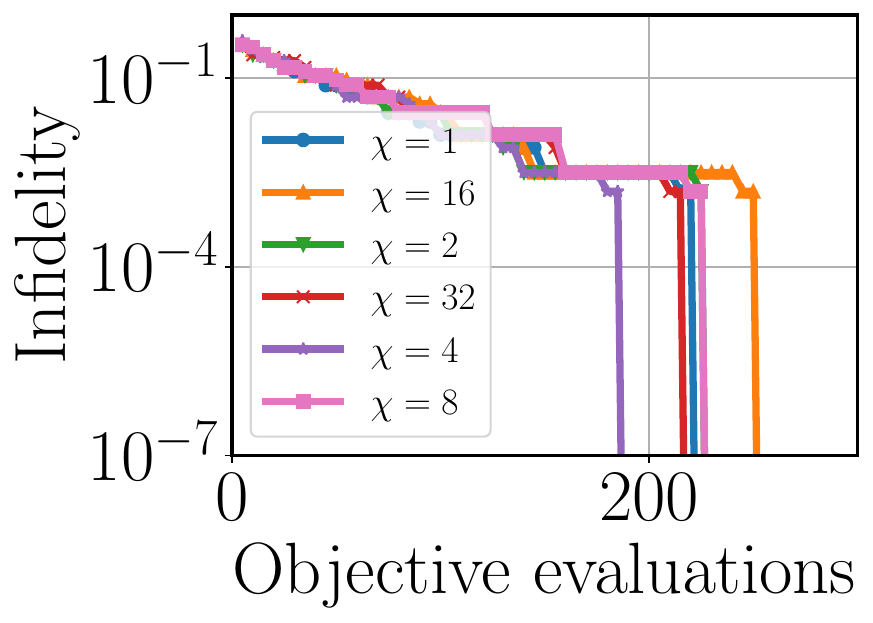}
        \label{fig:det0-bonddim}
    \end{subfigure}
    \hfill
    \begin{subfigure}{0.48\columnwidth}
        \centering
        \caption{}
        \includegraphics[width=\linewidth]{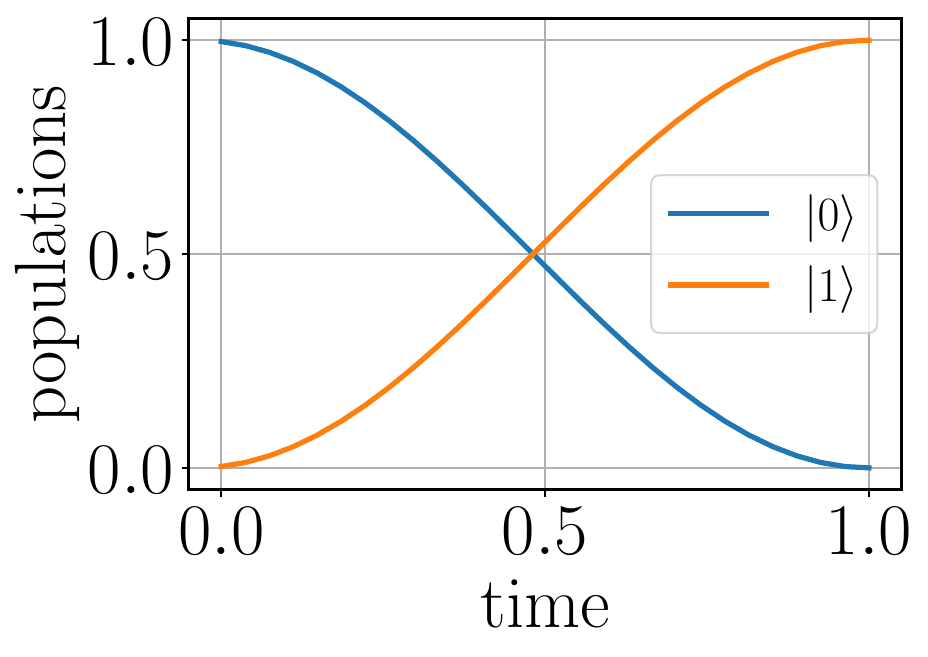}
        \label{fig:det0-pop}
    \end{subfigure}
    \caption{Single-qubit optimal control for population transfer \(\ket{0}\!\to\!\ket{1}\) with \(\Delta = 0\) is shown. System parameters are total time  \(T = \pi\), length of time-series (also the length of the MPS) \(L = 28\). Optimization parameters are accessible amplitude levels \(d = 2\), bond dimension \(\chi=4\), number of sample points \(K = 12\), number of elites \(k = 2\), learning rate \(\eta = 0.07\), and number of single-site sweeps 10. (a) Median convergence of the infidelity over 20 independent runs (different random seeds) is shown. The solid curve shows the median, the dashed curves show the 16th and 84th percentiles, and the shaded band spans this percentile range. (b) Comparison of 20 independent runs with varying bond dimensions is shown. (c) The corresponding population transfer dynamics is depicted. }
    \label{fig:singlequbit-det0}
\end{figure}

In the resonant \(\Delta = 0\) case, the optimal control corresponds to a constant-amplitude pulse that drives Rabi oscillations between the two states. The optimal pulse duration is given by \( T = \pi/ u_0 \). Figure~\ref{fig:singlequbit-det0} summarizes the results obtained using TT-EDA for this setting. The median convergence curve shown in Fig.~\ref{fig:det0-convergence} indicates that the method approaches high-fidelity solutions within a few hundred objective evaluations. Across independent runs, the convergence behavior exhibits relatively limited variability, as reflected by the shaded percentile region in Fig.~\ref{fig:det0-convergence}. When compared to the gradient-free baselines considered, TT-EDA reaches high-fidelity solutions within fewer objective evaluations than the gradient-free optimizers included in the comparison for this task. Among the methods tested, CMA exhibits convergence behavior that is closely comparable to TT-EDA under the same evaluation budget with both approaches approaching similar final infidelities. In Fig.~\ref{fig:det0-bonddim}, we analyze the hyperparameter sensitivity of TT-EDA with bond dimension. The results indicate that relatively small bond dimensions (e.g., \(\chi = 4\)) can be sufficient in this setting. The variance across different runs with varying bond dimensions is also small. The corresponding population dynamics shown in Fig.~\ref{fig:det0-pop} illustrate that the resulting control fields drive the system from \(\ket{0}\) to \(\ket{1}\) with high final-state fidelity.

\begin{figure}[t]
    \centering

    \begin{subfigure}{\columnwidth}
        \centering
        \caption{}
        \includegraphics[width=\linewidth]{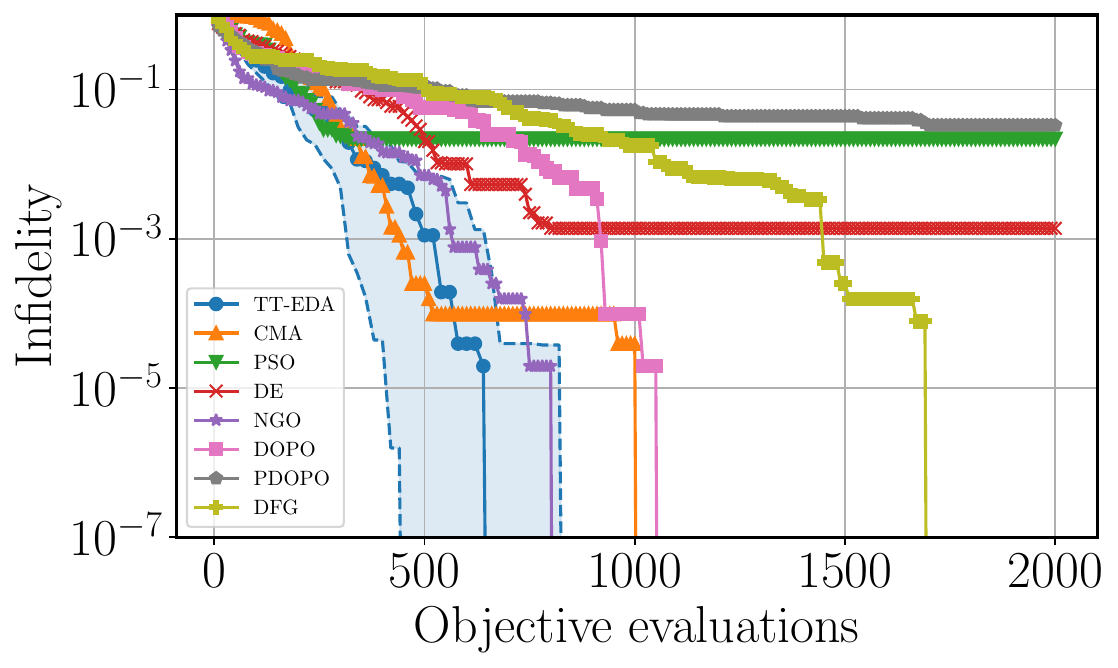}
        \label{fig:det1-convergence}
    \end{subfigure}
    \begin{subfigure}{0.48\columnwidth}
        \centering
        \caption{}
        \includegraphics[width=\linewidth]{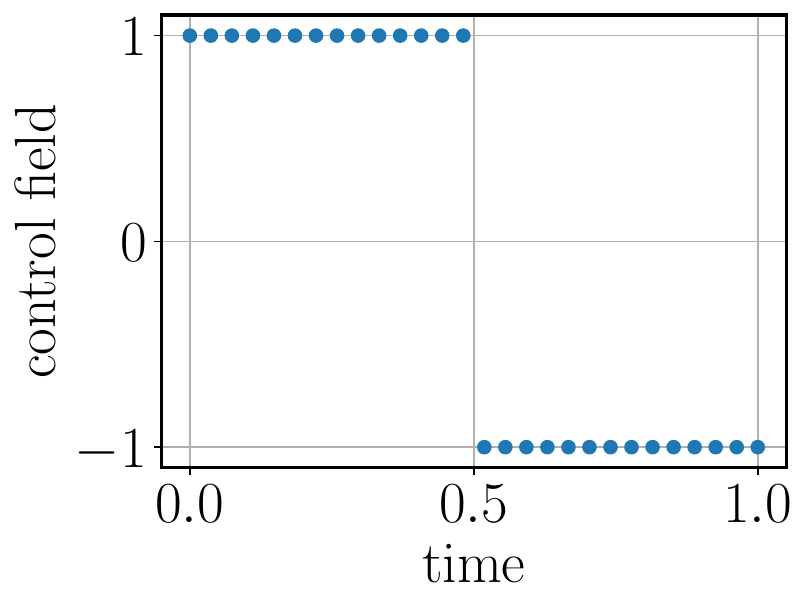}
        \label{fig:det1-laser}
    \end{subfigure}
 \hfill
    \begin{subfigure}{0.48\columnwidth}
        \centering
        \caption{}
        \includegraphics[width=\linewidth]{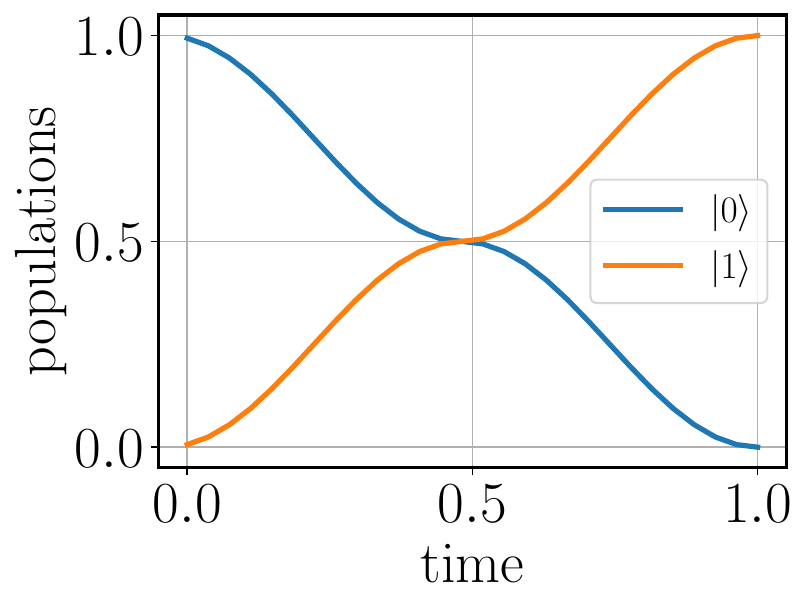}
        \label{fig:det1-pop}
    \end{subfigure}
    \caption{Single-qubit bang-bang optimal control for population transfer \(\ket{0}\!\to\!\ket{1}\) with \(\Delta = 1\) is shown. System parameters are set as total time \(T = \pi \sqrt{2}\), length of time-series \(L = 28\) time steps. Optimization parameters are accessible amplitude levels \(d = 3\), bond dimension \(\chi=5\), number of sample points \(K = 20\), number of elites \(k = 2\), learning rate \(\eta = 0.06\), and number of single-site sweeps 20. (a) Median convergence of the infidelity over 20 independent runs is given. The solid curve shows the median, the dashed curves show the 16th and 84th percentiles, and the shaded band spans this percentile range. (b) Optimized pulse profile exhibiting bang-bang structure and (c) corresponding population transfer are shown.}
    \label{fig:singlequbit-det1}
\end{figure}

In the off-resonant \(\Delta \neq 0\) case, the optimal solution is known to be bang-bang sequence where the control switches abruptly between its maximum and minimum values. In this setting, we examine whether the optimization procedure is able to recover this qualitative structure without explicitly enforcing it. Figure~\ref{fig:singlequbit-det1} summarizes the off-resonant results obtained by TT-EDA. The convergence behavior in Fig.~\ref{fig:det1-convergence} shows that TT-EDA approaches high-fidelity bang-bang solution within a few hundred objective evaluations. Within the evaluation budget considered, the method reaches the target fidelity in fewer evaluations than the other gradient-free optimizers included in the comparison. The optimized control profile displayed in Fig.~\ref{fig:det1-laser} exhibits a bang-bang structure consistent with the known solution. The associated population dynamics shown in Fig.~\ref{fig:det1-pop} illustrate that the resulting control field drives the system from \(\ket{0}\) to \(\ket{1}\) with high final-state fidelity.

\subsection{Bell-pair Preparation}
As a second benchmark, we consider the preparation of a maximally entangled Bell state
in a system of two coupled spin-$\tfrac{1}{2}$ particles driven by a global time-dependent magnetic field and interacting via an Ising coupling. This minimal composite system constitutes a canonical model for entanglement generation and has been extensively studied in quantum control literature. The model Hamiltonian is given by
\begin{equation}
\label{eq: bell-pair-hamiltonian}
H(t)
=
4\xi\, S_{1z}\otimes S_{2z}
+
\mu\,\mathbf{B}(t)\cdot
\left(
\mathbf{S}_1\otimes \mathbb{1}
+
\mathbb{1}\otimes \mathbf{S}_2
\right),
\end{equation}
where $\xi>0$ denotes the Ising coupling strength, $\mu$ is the gyromagnetic ratio,
$S_{i\alpha}$ $(\alpha=x,y,z)$ are spin operators, and $\mathbf{B}(t)$ is a global
time-dependent magnetic field acting identically on both spins. Using the standard triplet basis set and starting from the initial state $\ket{\downarrow\downarrow}$,
the dynamics remain confined to the triplet subspace spanned by \(\{\ket{\downarrow\downarrow},\ket{\downarrow \uparrow}_+, \ket{\uparrow\uparrow}\} \) with \(\ket{\downarrow \uparrow}_+=(\ket{\downarrow\uparrow}+\ket{\uparrow\downarrow})/\sqrt{2}\) while the singlet state \((\ket{\downarrow\uparrow}-\ket{\uparrow\downarrow})/\sqrt{2}\) is dynamically decoupled. Opting for a rotating transverse magnetic field and transforming to a rotating frame with respect to the carrier frequency, the rapidly oscillating terms are neglected within the rotating-wave approximation. The resulting effective Hamiltonian takes the form
\begin{equation}
    \label{eq: bellpair-effective-hamiltonian}
H_{\mathrm{eff}}(t) =
\begin{pmatrix}
\Delta(t) & \Omega(t)/\sqrt{2} & 0 \\
\Omega(t)/\sqrt{2} & 0 & \Omega(t)/\sqrt{2} \\
0 & \Omega(t)/\sqrt{2} & 4\xi - \Delta(t)
\end{pmatrix},
\end{equation}
expressed in the ordered basis
\(
\{\ket{\downarrow\downarrow}, \ket{\downarrow \uparrow}_+, \ket{\uparrow\uparrow}\}.
\)
Here, $\Omega(t)$ denotes the effective Rabi frequency and
\(\Delta(t)\)
is the detuning. This reduced three-level description captures all relevant dynamics for Bell-pair generation from the initial state $\ket{\downarrow\downarrow}$. The Rabi frequency $\Omega(t)$ controls population transfer between the triplet states while the detuning $\Delta(t)$ shifts their relative energies. In the following, this effective Hamiltonian is used directly for simulation. The goal is to transfer the system from the initial state \(\ket{\downarrow\downarrow}\) to the state \(\ket{\downarrow \uparrow}_+\) while minimizing the infidelity objective function \(\mathcal{J}=1 - \abs{{}_+ \braket{\downarrow \uparrow|\psi(T)}}^2\) at final time \(T\). 

\begin{figure}[!t]
    \centering
    \begin{subfigure}{\columnwidth}
        \centering
        \caption{}
        \includegraphics[width=\linewidth]{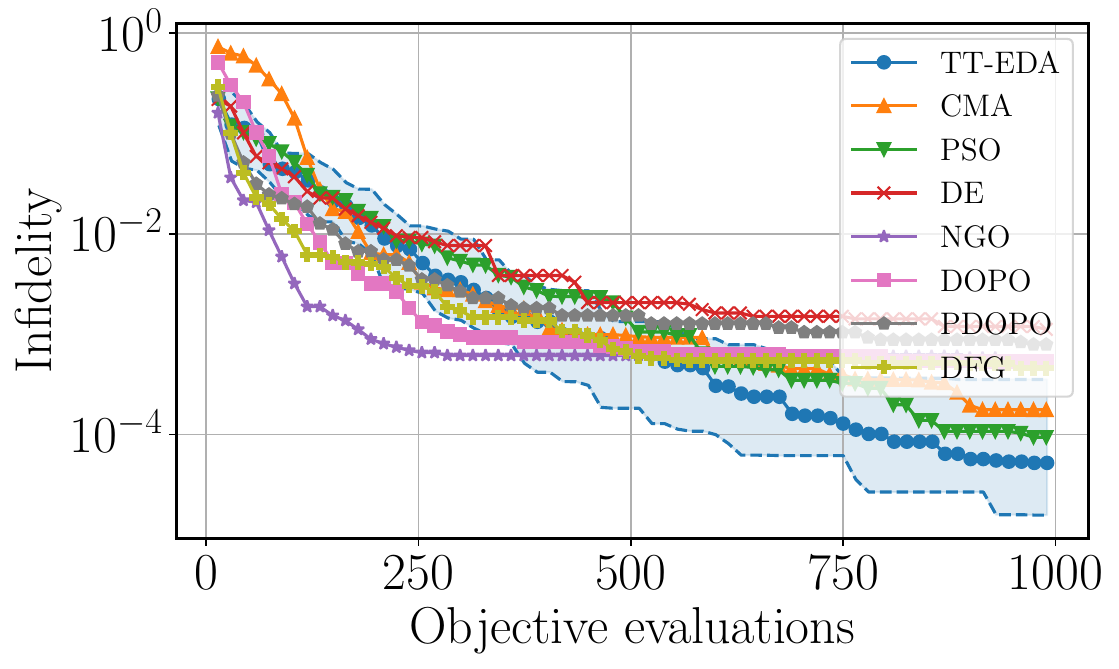}
        \label{fig:bell-convergence}
    \end{subfigure}
    \begin{subfigure}{0.48\columnwidth}
        \centering
        \caption{}
        \includegraphics[width=\linewidth]{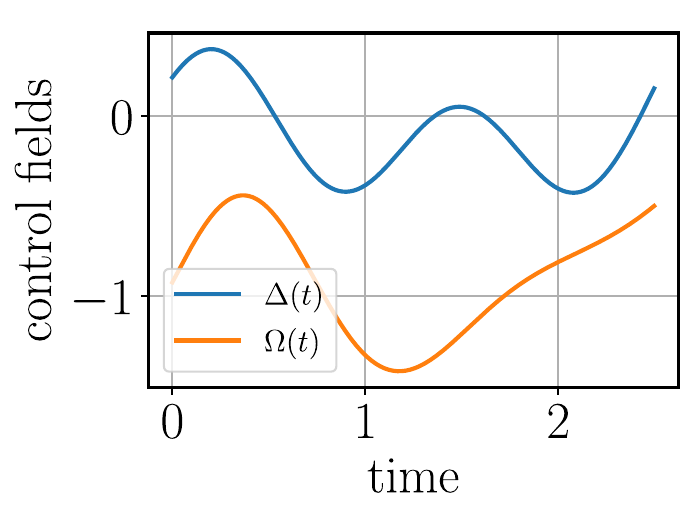}
        \label{fig:bell-laser}
    \end{subfigure}
    \hfill
    \begin{subfigure}{0.48\columnwidth}
        \centering
        \caption{}
        \includegraphics[width=\linewidth]{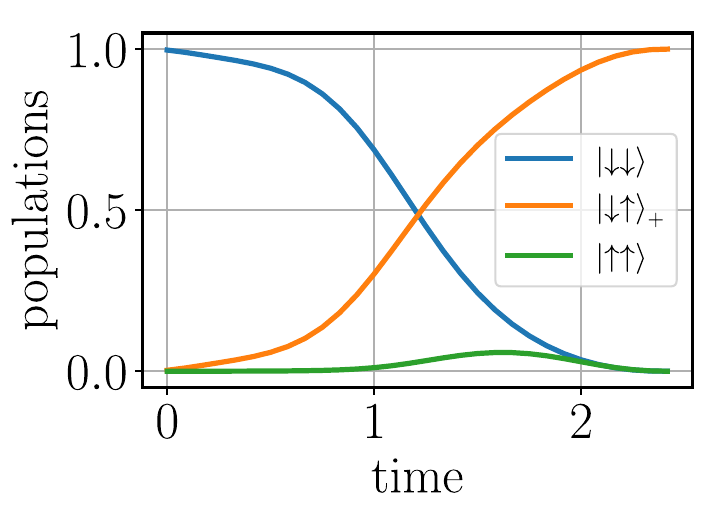}
        \label{fig:bell-pop}
    \end{subfigure}
    \caption{Optimal Bell-pair \(\ket{\downarrow\downarrow}\!\to\!\ket{\psi_+}\) preparation using Fourier basis encoding is shown. System parameters are total time set as \(T = 2.5\), number of time steps 30, Ising coupling strength \(\xi = 1\). Optimization parameters are number of Fourier basis \(J = 5\) for each Rabi frequency and detuning, bond dimension \(\chi=2\), accessible levels for each Fourier coefficients \(d = 40\), number of sample points \(K = 15\), number of elites \(k = 3\), learning rate \(\eta = 0.06\), and number of single-site sweeps 10. (a) Median convergence of the infidelity over 20 independent runs is given. The solid curve shows the median, the dashed curves show the 16th and 84th percentiles, and the shaded band spans this percentile range. (b) Optimized control fields Rabi frequency \(\Omega(t)\) and detuning \(\Delta(t)\) and (c) corresponding population dynamics are shown.}
    \label{fig:bellpair}
\end{figure}

In this benchmark, we employ a Fourier basis encoding for the control fields. We expand both the Rabi frequency \(\Omega(t)\) and detuning \(\Delta(t)\) in truncated Fourier series. The MPS score function is defined over the space of these Fourier coefficients. In Figure \ref{fig:bellpair}, we present the optimization results for Bell-pair generation using Fourier-encoded control fields. Figure \ref{fig:bell-convergence} shows the median convergence curve which demonstrates that TT-EDA achieves high-fidelity Bell-pair preparation with fewer function evaluations compared to baseline methods. In terms of run variance, the TT-EDA exhibits low variability across independent runs with low infidelities as it can be seen in Figure \ref{fig:bell-convergence}. Within this benchmark, TT-EDA, CMA, and PSO exhibit broadly similar convergence behavior under the fixed evaluation budget considered. While all three methods approach comparable final fidelities, TT-EDA is observed to attain the lowest infidelity among the tested approaches. In contrast, the discrete gradient-free optimizers included in the comparison do not achieve comparable performance within the same evaluation limit. In Fig.~\ref{fig:bell-laser}, the optimized control fields \(\Omega(t)\) and \(\Delta(t)\) are shown. The control fields exhibit smooth profiles characteristic of Fourier basis representations. The corresponding population dynamics in Fig.~\ref{fig:bell-pop} illustrate that the system evolves from the initial state \(\ket{\downarrow\downarrow}\) toward the target state \(\ket{\downarrow \uparrow}_+\) while population of the state \(\ket{\uparrow\uparrow}\) is small throughout the evolution.

\subsection{Single Qubit Gate in a Qutrit System}
In this case, we consider the implementation of a single-qubit gate in the
presence of an additional, weakly detuned energy level.
Such a setting naturally arises in many physical qubit platforms, including
superconducting circuits and nitrogen-vacancy centers, where the computational
subspace is embedded in a multilevel system. While the qubit is encoded in the two lowest levels, the proximity of higher-lying states can lead to leakage and gate errors if not properly accounted for. The system is modeled as a three-level ladder with weak anharmonicity. After moving to a rotating frame and applying the rotating-wave approximation, the dynamics are governed by the Hamiltonian
\begin{align}
\label{eq:qutrit-hamiltonian}
H(t)
&=
\Delta \ket{2}\!\bra{2}
+
\delta(t)\ket{1}\!\bra{1} \nonumber
\\ &+
\sum_{n=1}^{2}
\frac{\sqrt{n}}{2}
\Bigl[
c_x(t)\,\sigma^{x}_{n-1,n}
+
c_y(t)\,\sigma^{y}_{n-1,n}
\Bigr],
\end{align}
where
\(
\sigma^{x}_{n,m} = \ket{n}\!\bra{m} + \ket{m}\!\bra{n}
\)
and
\(
\sigma^{y}_{n,m} = i(\ket{m}\!\bra{n} - \ket{n}\!\bra{m})
\)
denote generalized Pauli operators acting between neighboring levels.
The factors $\sqrt{n}$ encode the ladder-type coupling strengths between adjacent
transitions, while $\delta(t)$ represents a detuning of the applied drive and \(\Delta\) is the anharmonicity. In the following, the detuning is fixed to $\delta(t)=0$, and control is exerted solely
via $c_x(t)$ and $c_y(t)$. The control objective is to implement a NOT gate on the computational subspace
spanned by $\{\ket{0},\ket{1}\}$, while minimizing population leakage into the level $\ket{2}$.
Gate performance is quantified by the average gate infidelity \(\mathcal{J}
=
1 - \frac{1}{6}
\sum_{j \in \{\pm x, \pm y, \pm z\}}
\bigl|
\bra{j} U_{\mathrm{t}}^{\dagger} U(T) \ket{j}
\bigr|^2\) where $U(T)$ is the propagator generated by Hamiltonian in Eq.~\ref{eq:qutrit-hamiltonian} and
$U_{\mathrm{t}}$ denotes the target NOT gate embedded in the three-level Hilbert space.

\begin{figure}[!t]
    \centering
    \begin{subfigure}{\columnwidth}
        \centering
        \caption{}
        \includegraphics[width=\linewidth]{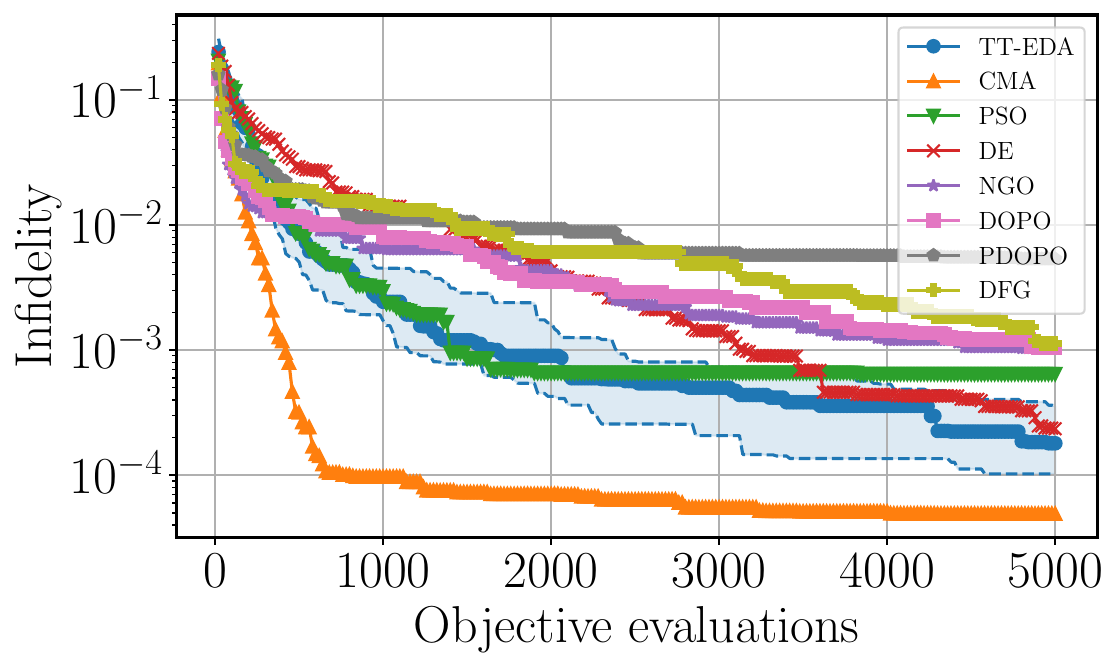}
        \label{fig:qutrit-avg}
    \end{subfigure}
    \begin{subfigure}{0.48\columnwidth}
        \centering
        \caption{}
        \includegraphics[width=\linewidth]{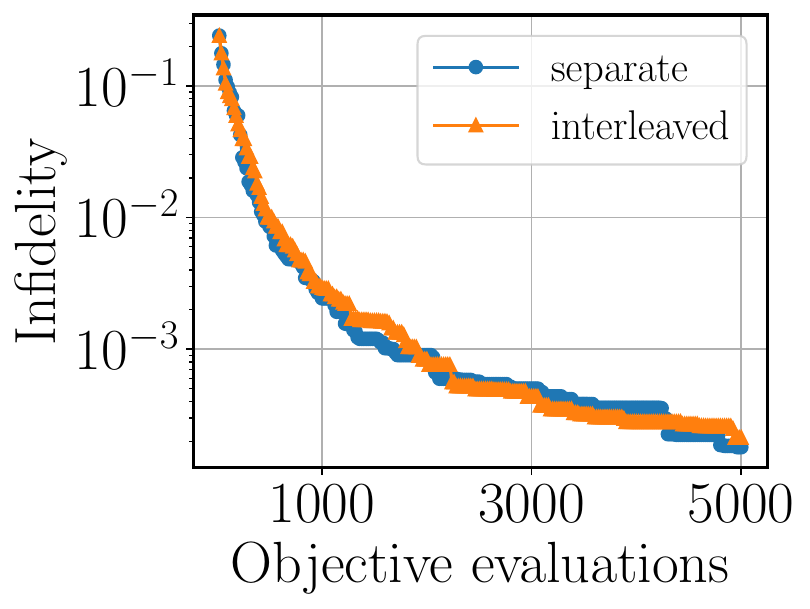}
        \label{fig:qutrit-interleaved}
    \end{subfigure}
    \hfill
    \begin{subfigure}{0.48\columnwidth}
        \centering
        \caption{}
        \includegraphics[width=\linewidth]{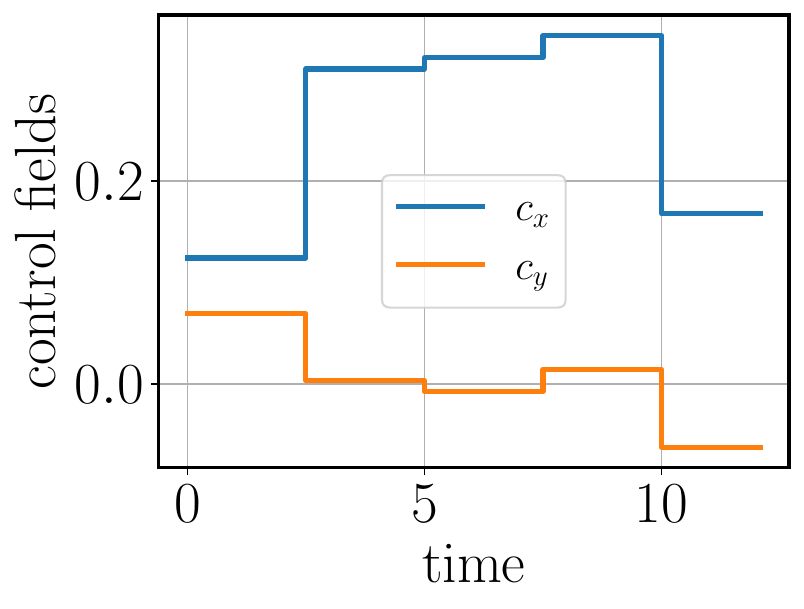}
        \label{fig:qutrit-laser}
    \end{subfigure}
    \caption{Qutrit NOT-gate optimization using piecewise-constant control fields. Parameters are given as anharmonicity \( \Delta = -400 \times 2\pi ~\mathrm{MHz} \), total time \( T = 12.5/|\Delta| \), number of piecewise-constant basis \( J = 5 \) with for each quadrature, bond dimension \(\chi=4\), accessible levels for each coefficient \(d = 50\), number of sample points \(K = 20\), number of elites \(k = 2\), learning rate \(\eta = 0.07\), and number of single-site sweeps 10. (a) Median convergence of the NOT-gate infidelity over 20 independent runs. The solid curve shows the median, the dashed curves show the 16th and 84th percentiles, and the shaded band spans this percentile range. (b) Performance comparison of interleaved versus separate ordering of the MPS tensors is displayed. (c) Example of the optimized control fields is depicted. Both the horizontal (time) and vertical (amplitude) axes are in units of \(\abs{\Delta}\).}
    \label{fig:qutrit}
\end{figure}

In this test case, we employ a piecewise-constant basis parameterization for the control fields $c_x(t)$ and $c_y(t)$. The MPS score function is defined over the space of these discrete control amplitudes at each piecewise-constant segment. Figure \ref{fig:qutrit} shows the results for qutrit NOT-gate optimization. The median convergence behavior shown in Fig.~\ref{fig:qutrit-avg} indicates that TT-EDA, PSO, and DE exhibit broadly similar convergence trends under the fixed evaluation budget considered. Among these methods, TT-EDA is observed to reach the lowest final gate infidelity within the allotted number of function evaluations. CMA-ES attains the lowest infidelity overall in this benchmark and converges rapidly to a high-fidelity solution within a few thousand evaluations. While TT-EDA displays slower progress during the early stages of the optimization, it continues to improve with additional evaluations and approaches comparable fidelity levels later in the run. The variability across independent runs for TT-EDA is relatively limited, as reflected in Fig.~\ref{fig:qutrit-avg}. In this test case, we also examine the influence of tensor ordering in the MPS representation. Specifically, we compare an interleaved ordering where tensors associated with the two quadratures are alternated and with a separate ordering in which all tensors corresponding to one quadrature precede those of the other. The comparison shown in Fig.~\ref{fig:qutrit-interleaved} indicates that both orderings lead to similar convergence behavior and final infidelities across multiple runs. The corresponding optimized control fields are shown in Fig.~\ref{fig:qutrit-laser}.

\subsection{Population Transfer in a Three-Level Open System}
As the last benchmark, we consider population transfer in a three-level system for which Stimulated Raman Adiabatic Passage (STIRAP) provides a solution. Beyond its fundamental relevance, this system provides a natural testbed for optimal control particularly in the presence of dissipation. We consider a three-level system with states
$\{\ket{g}, \ket{e}, \ket{r}\}$. The transitions $\ket{g}\leftrightarrow\ket{e}$ and $\ket{e}\leftrightarrow\ket{r}$ are driven
by two classical control fields (pump and Stokes) with time-dependent Rabi frequencies
$\Omega_p(t)$ and $\Omega_s(t)$, respectively.
In a suitable rotating frame and within the rotating-wave approximation, the Hamiltonian reads
\begin{align}
H(t)
&=
\Delta_p \ket{e}\!\bra{e}
+
\Delta_3 \ket{r}\!\bra{r}
+
\frac{\Omega_p(t)}{2}
\left(
\ket{g}\!\bra{e} + \ket{e}\!\bra{g}
\right) \nonumber \\
&+
\frac{\Omega_s(t)}{2}
\left(
\ket{e}\!\bra{r} + \ket{r}\!\bra{e}
\right),
\label{eq:stirap_hamiltonian}
\end{align}
where $\Delta_p$ is the single-photon detuning and $\Delta_3=\Delta_p-\Delta_s$ is the
two-photon detuning, which is set to zero in the following. To account for losses from the excited state $\ket{e}$, the system dynamics are
described by a Lindblad master equation
\begin{equation}
\dot{\rho}(t)
=
-i[H(t),\rho(t)]
+
\mathcal{L}_\gamma[\rho(t)],
\label{eq:stirap-master-equation}
\end{equation}
with
\begin{equation}
\label{eq:Lindbladian}
\mathcal{L}_\gamma[\rho]
=
\frac{\gamma}{2}
\left(
2\ket{s}\!\bra{e}\rho\ket{e}\!\bra{s}
-
\ket{e}\!\bra{e}\rho
-
\rho\ket{e}\!\bra{e}
\right),
\end{equation}
where $\gamma$ is the decay rate and $\ket{s}$ is an auxiliary sink state collecting population losses outside the three-level manifold. The control task is a population transfer from state \(\ket{g}\) to state \(\ket{r}\) with low infidelity despite the presence of dissipation where \(\ket{e}\) population should be suppressed. The STIRAP protocol achieves this by adiabatically following a dark state that avoids populating the lossy intermediate state \(\ket{e}\). The counterintuitive pulse sequence, where the Stokes pulse precedes the pump pulse, enables efficient population transfer from \(\ket{g}\) to \(\ket{r}\) without significant occupation of \(\ket{e}\). In this test case, we investigate whether the TT-EDA rediscovers this known optimal control strategy. Starting from the initial state $\rho(0)=\ket{g}\!\bra{g}$, the objective is to maximize the population of the target state $\ket{r}$ at a fixed final time $T$. The figure of merit is defined as \(\mathcal{J}=1 - \mathrm{Tr}\!\left[\rho(T)\ket{r}\!\bra{r}\right] \). 

\begin{figure}[!t]
    \centering
    \begin{subfigure}{\columnwidth}
        \centering
        \caption{}
        \includegraphics[width=\linewidth]{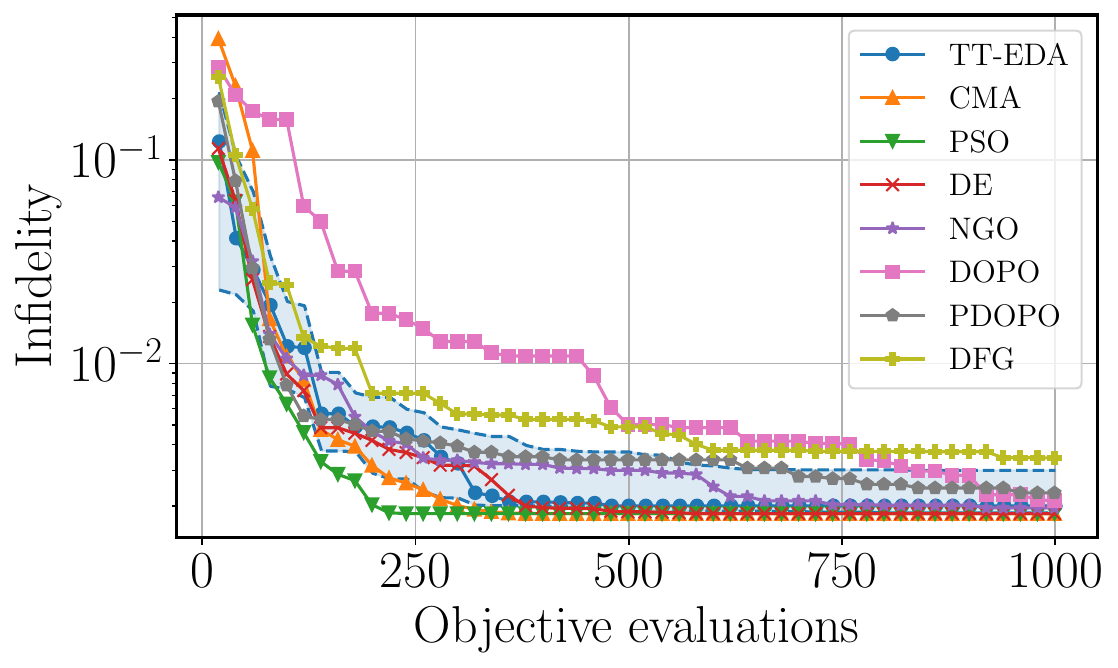}
        \label{fig:stirap-conv}
    \end{subfigure}
    \begin{subfigure}{0.48\columnwidth}
        \centering
        \caption{}
        \includegraphics[width=\linewidth]{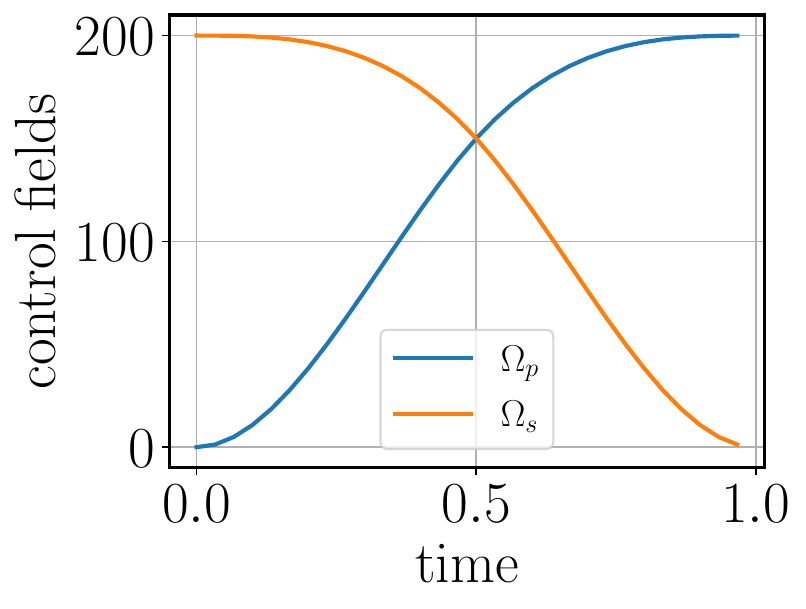}
        \label{fig:stirap-pulse}
    \end{subfigure}
    \hfill
    \begin{subfigure}{0.48\columnwidth}
        \centering
        \caption{}
        \includegraphics[width=\linewidth]{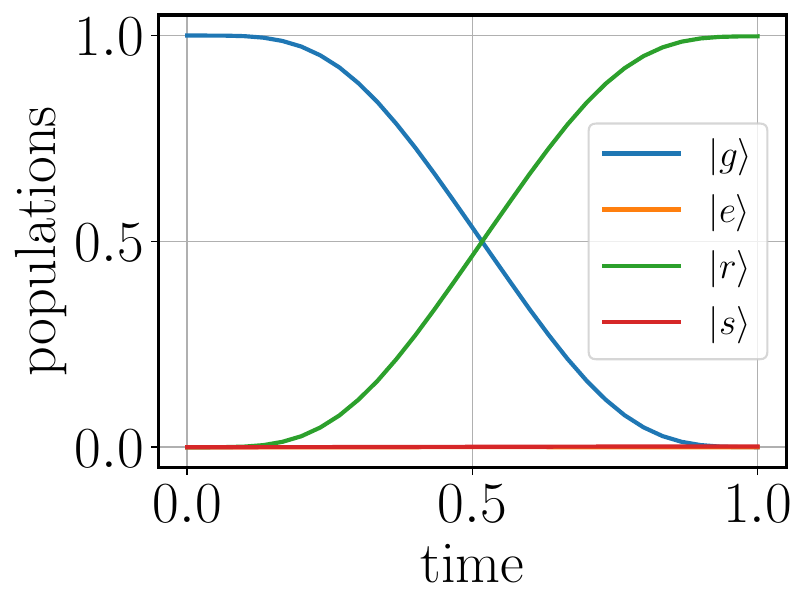}
        \label{fig:stirap-pop}
    \end{subfigure}
    \caption{Optimal population transfer in a three-level open system using spline basis encoding is shown. Parameters are set as total time set as \(T = 1\), with total number of time steps 30, number of spline coefficients \(J = 10\) for both pump and Stokes Rabi frequencies, bond dimension \(\chi=5\), and decay rate \(\gamma = 5\). Optimization parameters are accessible levels for each spline coefficient \(d = 10\), number of sample points \(K = 20\), number of elites \(k = 2\), learning rate \(\eta = 0.06\), and number of single-site sweeps 10. (a) Median convergence of the infidelity over 20 independent runs is given. The solid curve shows the median, the dashed curves show the 16th and 84th percentiles, and the shaded band spans this percentile range. (b) Optimized control fields: pump \(\Omega_p(t)\) and Stokes \(\Omega_s(t)\) Rabi frequencies. (c) Corresponding population dynamics.}
    \label{fig:stirap}
\end{figure}

In this test case, we employ a spline basis encoding for the control fields. We 
expand both Rabi frequencies $\Omega_p(t)$ and $\Omega_s(t)$ in B-spline bases of degree \( p=3 \).
The MPS score function is defined over the space of the spline coefficients. Figure~\ref{fig:stirap} summarizes the optimization results for the open-system population transfer benchmark. The median convergence behavior shown in Fig.~\ref{fig:stirap-conv} indicates that TT-EDA, CMA, PSO, and DE exhibit broadly similar convergence trends during the early stages of the optimization, reaching high-quality solutions within a relatively small number of objective evaluations compared to the other methods considered. Over this initial evaluation regime, these approaches display comparable rates of improvement, while the remaining optimizers approach similar fidelity levels only at later stages of the optimization. Across the full evaluation budget, TT-EDA continues to refine the solution and attains final infidelities that are comparable to those obtained by the strongest continuous gradient-free baselines. The variability across independent runs for TT-EDA remains limited, as reflected by the shaded percentile region in Fig.~\ref{fig:stirap-conv}. Continuous optimizers such as CMA-ES, PSO, and DE therefore remain strong reference methods for this benchmark, which is consistent with their established performance in smooth optimization settings. The optimized control fields obtained with TT-EDA exhibit the counterintuitive pulse ordering characteristic of STIRAP, as shown in Fig.~\ref{fig:stirap-pulse}, with the Stokes pulse \(\Omega_s(t)\) preceding the pump pulse \(\Omega_p(t)\). The corresponding population dynamics in Fig.~\ref{fig:stirap-pop} illustrate that the system evolves from the initial state \(\ket{g}\) to the target state \(\ket{r}\) while maintaining a low population of the lossy intermediate state \(\ket{e}\) throughout the evolution in the presence of dissipation.
    
\section{Conclusion and Outlook}
\label{seq:outlook}
In this work, we introduced a tensor network adaptive sampling heuristic for quantum optimal control and presented a numerical study of its performance. Specifically, by representing score function over the search space of control parameters using matrix product states, the method provides a structured way to locate optimal control parameters in the control landscape. We assessed its behavior across a range of representative quantum control problems, including single-qubit state transfer, Bell-pair preparation, qutrit gate implementation, and population transfer in open quantum systems. Across these benchmarks, we observed that the approach can achieve performance competitive to established gradient-free optimization methods in certain settings, while exhibiting different convergence characteristics depending on the problem structure and control encoding. 

In the single-qubit control tasks considered, we observed that the method is able to recover pulse shapes associated with high-fidelity state transfer, including in the presence of detuning. In the Bell-pair generation benchmark, our approach was found to identify control solutions leading to entangled target states within a moderate number of function evaluations. The qutrit gate example illustrates that the method can be applied to multilevel systems where leakage effects are relevant, while the open-system population transfer task provides a setting to examine its behavior in the presence of dissipation. Across the benchmarks studied, TT-EDA generally exhibits performance comparable to commonly used discrete gradient-free optimizers when evaluated under equal function-evaluation budgets. Continuous optimizers such as CMA-ES remain strong and competitive baselines, and in several cases we observe comparable performance between TT-EDA and CMA-ES depending on the problem and encoding. Overall, these results suggest that tensor network parameterizations can provide a viable representation for exploring structured quantum control landscapes, while leaving open questions regarding their relative advantages across different regimes.

This study also highlights several avenues that merit further investigation. The discrete encoding of control parameters implies that the optimization is performed over an approximate control space, which may introduce discretization-related errors. While finer discretizations could in principle improve the achievable fidelity, this would come at increased computational cost. In addition, the present implementation does not make explicit use of continuity information in the control landscape, which may influence optimization efficiency in certain settings. We also observe that the optimization of the tensor network parameters can, in some cases, exhibit signs of premature convergence, potentially limiting exploration of the control space. Finally, the choice of hyperparameters such as bond dimension, learning rate, and mutation strength was guided by empirical considerations, and the development of more systematic tuning strategies remains an open direction for future work. Taken together, these suggest that TT-EDA, in its current form, is most naturally aligned with quantum control problems that are inherently discrete or strongly discretized while its application to fully continuous control landscapes may benefit from hybrid or complementary strategies.

Looking ahead, this study suggests several directions that could be explored in future work. On the methodological side, possible extensions include the use of alternative tensor network architectures, modifications to the update scheme, and the incorporation of continuous control parameterizations. The integration of additional sources of information beyond function evaluations such as local optimization information could also be examined as a way to influence optimization behavior. Further work may also consider approaches for handling constraints more explicitly within the optimization framework. From an application perspective, it would be of interest to investigate the behavior of the proposed method in larger quantum systems, in the presence of noise and model uncertainty. Control problems with inherently discrete structure, such as certain quantum circuit design tasks, provide a natural context in which to further assess the suitability of these methods. An additional potential use of the method can be a source of structured initial conditions for gradient-based quantum control methods. By moderating the concentration of the scoring through less aggressive updates or increased exploration the method may be used to sample control configurations that already achieve moderate fidelity and capture relevant structure of the control landscape. Such configurations could serve as initial guesses for subsequent gradient-based refinement. Exploring these directions would help clarify the scope and limitations of the method in quantum control.
\begin{acknowledgments}
This work is funded by the Cluster of Excellence ``CUI: Advanced Imaging of Matter'' of the Deutsche Forschungsgemeinschaft (DFG) - EXC 2056 - Project ID 390715994. Z. Zeybek acknowledges suggestions from Donika Imeri on the illustrative depictions.
\end{acknowledgments}

\appendix
\begin{widetext}
\section{Comparison and Connection with PROTES}
\label{app:protes}
In this section, we provide additional material on TT-EDA and discuss as to how it compares and related with PROTES \cite{PROTES2023} and show benchmark results. The non-negative TT/MPS represents a score function \(S_{\theta}(\mathbf{x})\) over the search space of discretized control parameters as follows,
\begin{equation}
\label{eq:mps-score}
S_\theta(\mathbf{x}) =
\sum_{\alpha_1,\ldots,\alpha_{L-1}}
A^{[1]}_{1,\alpha_1}(x_1)\,
A^{[2]}_{\alpha_1,\alpha_2}(x_2)\cdots
A^{[L]}_{\alpha_{L-1},1}(x_L).
\end{equation}
The whole utility of using the MPS representation is that $1)$ it provides an efficient way to represent and manipulate score function over high-dimensional discrete spaces that would otherwise be infeasible to handle directly due to the curse of dimensionality $2)$ efficient generation of new better candidate solutions by sequential sampling from the score induced probability distribution. In this way, as long as we can use this score function to discriminate good control configurations from bad ones, we can use it to guide the search for optimal control parameters. Therefore, a general heuristic can be considered as iteratively updating the MPS parameters \(\theta\) such that the score function \(S_\theta(\mathbf{x})\) assigns higher values to better control configurations as rapidly as possible while maintaining diversity in the search. Conceptually, this can be represented as combination of exploit and explore terms. This can be mathmematically formulated as follows,
\begin{equation}
\label{eq:score-optimization}
\max_\theta \mathcal{L(\theta)}=\max_\theta
\frac{1}{M}\sum_{i=1}^M
\log S_\theta(\mathbf{x}^{i})
- \lambda \mathcal{R}(\theta),
\end{equation}
where \(\{\mathbf{x}^{i}\}_{i=1}^M\) are the elite samples drawn from the current score function \(S_\theta(\mathbf{x})\), \(\lambda > 0\) is a regularization parameter, and \(\mathcal{R}(\theta)\) is a regularization term that encourages exploration by penalizing overly concentrated score functions. The first term in the above equation encourages the score function to assign higher values to elite samples, while the second term discourages the score function from becoming too peaked around a small set of configurations. In TT-EDA, we do not include an explicit regularization term and only consider the first term in the above equation. This leads to a simpler optimization problem that focuses solely on increasing the scores of the elite samples. To see this more clearly, we can consider the induced probability distribution over the discrete control parameters as,
\begin{equation}
\label{eq:induced-probability}
P_\theta(\mathbf{x}) =
\frac{S_\theta(\mathbf{x})}
{Z_{\theta}}, \quad Z_\theta = \sum_{\mathbf{x}} S_\theta(\mathbf{x}),
\end{equation}
where the summation in the partition function \(Z_\theta\) runs over all possible discrete control configurations \(\mathbf{x}\). If we maximize the log-likelihood of the \(M\) elite samples \(\{\mathbf{x}^{i}\}_{i=1}^M\) drawn from the current distribution \(P_\theta(\mathbf{x})\), we obtain the following optimization problem,
\begin{equation}
\label{eq:objective-optimization}
\max_\theta \mathcal{L(\theta)}=\max_\theta
\frac{1}{M}\sum_{i=1}^M
\log P_\theta(\mathbf{x}^{i})
=
\max_\theta
\frac{1}{M}\sum_{i=1}^M
\log S_\theta(\mathbf{x}^{i})
-
\log Z_\theta.
\end{equation}
Using gradient ascent to optimize the above log-likelihood loss function leads to the following,
\begin{align}
\nabla_\theta \mathcal{L}(\theta)
&= \frac{1}{M}\sum_{i=1}^M
\nabla_\theta \log S_\theta(\mathbf{x}^{i}) - \nabla_\theta \log Z_\theta \nonumber\\
&= \frac{1}{M}\sum_{i=1}^M \frac{\nabla_{\theta}S_\theta(\mathbf{x}^{i})} {S_\theta(\mathbf{x}^{i})} - \frac{\nabla_{\theta}Z_\theta} {Z_\theta}.
\end{align}
Using the definition of the partition function \(Z_\theta\) in Eq.~\eqref{eq:induced-probability} and using \(\nabla_{\theta}S_\theta = S_{\theta}\nabla_{\theta}\log S_{\theta}\), the second term above can be rewritten as follows,
\begin{equation}
\frac{\nabla_{\theta}Z_\theta} {Z_\theta}
= \frac{1}{Z_\theta}\sum_{\mathbf{x}} \nabla_{\theta}S_\theta(\mathbf{x}) 
 = \frac{1}{Z_\theta} \sum_{\mathbf{x}} S_\theta(\mathbf{x}) \nabla_{\theta}\log S_\theta(\mathbf{x}) = \sum_{\mathbf{x}} \frac{S_\theta(\mathbf{x})}{Z_\theta} \nabla_{\theta}\log S_\theta(\mathbf{x}) = \sum_{\mathbf{x}} P_\theta(\mathbf{x}) \nabla_{\theta}\log S_\theta(\mathbf{x}) = \mathbb{E}_{\mathbf{x}\sim P_\theta}
\left[\nabla_\theta \log S_\theta(\mathbf{x})\right].
\end{equation}
where \(\mathbb{E}_{\mathbf{x}\sim P_\theta}[\nabla_\theta \log S_\theta(\mathbf{x})]\) denotes the expectation valueof the gradient of the log-score under the model distribution \(P_\theta(\mathbf{x})\). Therefore, the full gradient of the log-likelihood function can be expressed as
\begin{equation}
\nabla_\theta \mathcal{L}(\theta)
= \frac{1}{M}\sum_{i=1}^M \frac{\nabla_{\theta}S_\theta(\mathbf{x}^{i})} {S_\theta(\mathbf{x}^{i})} - \mathbb{E}_{\mathbf{x}\sim P_\theta}
\left[\nabla_\theta \log S_\theta(\mathbf{x})\right].
\end{equation}
So basically, the first term in the above equation is the empirical average of the gradient of the log-score over the elite samples, while the second term is the model average of the same quantity under the current distribution \(P_\theta(\mathbf{x})\). In this way, the resulting gradient balances attraction toward elite samples with a normalization term given by the model's own average log-score gradient. This enforces relative redistribution of probability mass rather than absolute score growth. In TT-EDA, we only consider the first term in the above equation and ignore the partition function term \(\log Z_\theta\) during the optimization, which leads to 
\begin{equation}
    \max_\theta \frac{1}{M}\sum_{i=1}^M \log S_\theta(\mathbf{x}^{i}),
\end{equation}
as given in the main text. This leads to score maximization where we push up the scores of the elite samples. Therefore, in our case, we only have a sampling distribution which is induced during the sampling step. Meaning that while the model implies a valid normalized probability distribution from \(P_{\theta}\) which we sample autoregressively, the optimization update ignores the global normalization constraint imposed by the partition function.

The intuition behind this approach is that \textit{optimization} does not necessarily require a proper density estimation. We rather focus on modifying the relative scores of the samples to favor elite samples more strongly in the next iteration. We are aware that ignoring the normalization term removes the model average that redistributes probability mass, which can cause rapid concentration onto a small subset of elites. This can consequently reduce diversity in the sampled configurations. To address this, a scaling parameter \(\lambda\) might be introduced in front of the partition function as,
\begin{equation}
\max_\theta \mathcal{L(\theta)}=\max_\theta
\frac{1}{M}\sum_{i=1}^M
\log S_\theta(\mathbf{x}^{i})
- \lambda \log Z_\theta.
\end{equation}
Conceptually, this can be then seen as a generalized update that balances pure probabilistic modeling with score maximization. Here, $\lambda$ controls the strength of the normalization term and affects the degree of concentration of the induced sampling distribution. In this way, TT-EDA would correspond to \(\lambda = 0\). In \(\lambda = 1\) case, we have the standard maximum likelihood estimation as in PROTES. Exploration and exploitation trade-off can then be tuned via this parameter. 

In the following, we provide an empirical comparison between the proposed TT-EDA approach and PROTES \cite{PROTES2023}. We turn of mutation in TT-EDA so that both methods rely on only tensor sampling and update steps. We evaluate both methods on a set of standard multivariate benchmark functions commonly used in the optimization literature and employ the original PROTES implementation provided by the authors \cite{PROTES2023}. Figure~\ref{fig:func_opt_compare} reports representative results for the Alpine, Ackley, Rastrigin, Griewank, Schwefel functions, all of which exhibit highly multimodal landscapes and pose nontrivial challenges for global optimization. A consistent scaling behavior across all benchmark functions can be seen in Figure~\ref{fig:func_opt_compare}. In low-dimensional settings (first row in Fig.~\ref{fig:func_opt_compare}), PROTES shows competitive or superior performance while TT-EDA occasionally converges prematurely. This is likely due to early concentration of sampling mass and reduced exploration. However, as the number of variables increases, TT-EDA consistently outperforms PROTES across all considered functions, both in terms of convergence speed and final objective value. Operationally, this crossover behavior reflects the differing update dynamics of the two methods. PROTES considers a probabilistic update that can admit a stable exploration. This can be beneficial in low-dimensional settings. In contrast, TT-EDA employs a more aggressive score-based update that concentrates probability mass toward elite regions more rapidly. This can lead to improved scalability in high-dimensional search spaces. Across this particular benchmark functions, these results indicate that TT-EDA exhibits more favorable scaling behavior with dimensionality within the budget considered here while PROTES remains more competitive in low-dimensional regimes. Both methods show low variance across runs. We acknowledge that this comparison is limited in scope and is empirical in nature. A more comprehensive theoretical analysis of the update dynamics and their implications for exploration-exploitation trade-offs would be valuable for understanding the observed behaviors. Especially, there is no guarantee that beyond the evaluation budget considered here the observed trends will persist. For instance, it might be the case that TT-EDA collapses early to suboptimal configurations but can only be made apparent in longer runs, including large varibale cases in Fig.~\ref{fig:func_opt_compare}. 
\begin{figure*}[t]
    \centering
    \includegraphics[width=\textwidth]{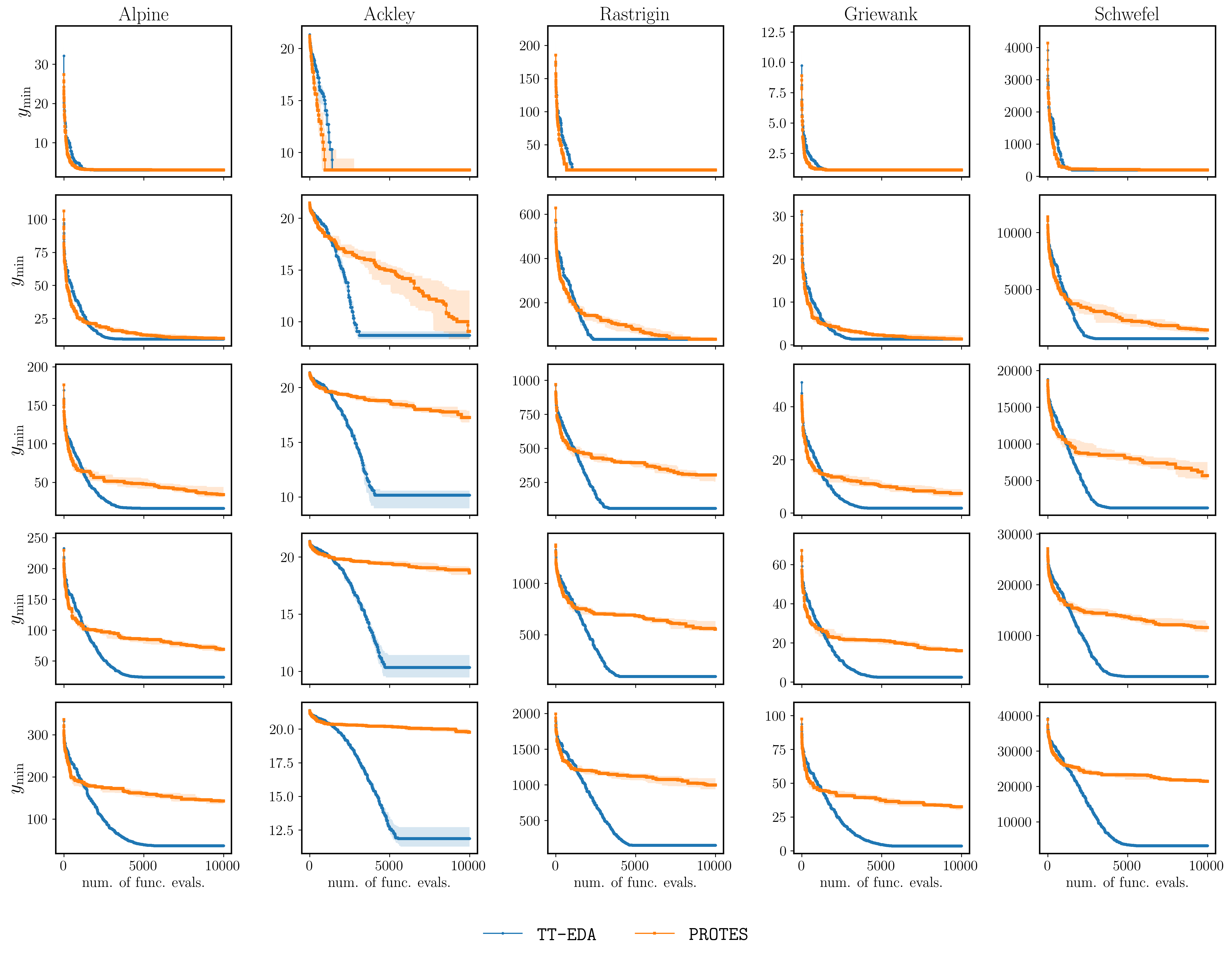}
    \caption{Convergence behavior of TT-EDA and PROTES across benchmark functions and problem dimensions.
Columns correspond to benchmark functions (Alpine, Ackley, Rastrigin, Griewank, Schwefel).
Rows correspond to the number of arguments (10, 30, 50, 70, 100). 
Each panel reports the median best objective value as a function of evaluations with interquantile ranges, computed over 10 independent runs. Local dimension \(d=16\) accessible values is used for each variable in all functions across all runs in both optimizers. Across all the runs the following configrations are used for both optimizers: bond dimension \(\chi = 4\), number of sample points \(K = 30\), number of elites \(M = 5\), learning rate \(\eta = 0.05\), and number of gradient updates 5.}
    \label{fig:func_opt_compare}
\end{figure*}

\end{widetext}
\bibliography{references.bib} 
\end{document}